\documentclass[twoside,twocolumn,9pt]{article}
\usepackage{extsizes}
\usepackage[super,sort&compress,comma]{natbib} 
\usepackage[version=3]{mhchem}
\usepackage[left=1.5cm, right=1.5cm, top=1.785cm, bottom=2.0cm]{geometry}
\usepackage{balance}
\usepackage{times,mathptmx}
\usepackage{sectsty}
\usepackage{graphicx} 
\usepackage{lastpage}
\usepackage[format=plain,justification=justified,singlelinecheck=false,font={stretch=1.125,small,sf},labelfont=bf,labelsep=space]{caption}
\usepackage{float}
\usepackage{fancyhdr}
\usepackage{fnpos}
\usepackage[english]{babel}
\addto{\captionsenglish}{%
  
}
\usepackage{array}
\usepackage{droidsans}
\usepackage{charter}
\usepackage[T1]{fontenc}
\usepackage[usenames,dvipsnames]{xcolor}
\usepackage{setspace}
\usepackage[compact]{titlesec}
\usepackage{amsmath}
\usepackage{bbold}
\usepackage{dsfont}
\usepackage{amssymb}
\usepackage[utf8]{inputenc}
\usepackage{graphicx}
\usepackage[colorlinks,allcolors=cyan!70!black]{hyperref}
\usepackage{lscape}
\usepackage{tabularx}
\usepackage{subfigure}
\usepackage{multicol}
\usepackage{stfloats}
\usepackage{enumitem}
\newcommand{\numberthis}{\addtocounter{equation}{1}\tag{\theequation}}
\newcolumntype{Y}{>{\centering\arraybackslash}X}
\newcommand{\vb}[1]{\mathbf{#1}} 
\newcommand{\rb}{\vb{r}}
\usepackage{bm}
\newcommand{\ten}[1]{\overset{\text{\tiny$\bm\leftrightarrow$}}{#1}}
\newcommand{\eq}{Eq. }
\newcommand{\fig}{Fig. }
\newcommand{\Fint}{\ensuremath{\vb{F}^\textrm{int}}}
\newcommand{\Fadh}{\ensuremath{\vb{F}^\textrm{adh}}}
\newcommand{\Fthresh}{\ensuremath{F^\textrm{thresh}}}
\newcommand{\Fstar}{\ensuremath{F^{\ast}}}
\newcommand{\xithresh}{\ensuremath{\xi^\textrm{thresh}}}
\newcommand{\Freq}{\ensuremath{\vb{F}^\textrm{req}}}
\newcommand{\beginsupplement}{
        \setcounter{table}{0}
        \renewcommand{\thetable}{S\arabic{table}}
        \setcounter{figure}{0}
        \renewcommand{\thefigure}{S\arabic{figure}}
        \setcounter{equation}{0}
        \renewcommand{\theequation}{S\arabic{equation}}
        \renewcommand{\thesection}{\arabic{section}}
     }

\usepackage{epstopdf}

\definecolor{cream}{RGB}{222,217,201}

\begin{document}

\pagestyle{fancy}
\thispagestyle{plain}
\fancypagestyle{plain}{
\renewcommand{\headrulewidth}{0pt}
}

\makeFNbottom
\makeatletter
\renewcommand\LARGE{\@setfontsize\LARGE{15pt}{17}}
\renewcommand\Large{\@setfontsize\Large{12pt}{14}}
\renewcommand\large{\@setfontsize\large{10pt}{12}}
\renewcommand\footnotesize{\@setfontsize\footnotesize{7pt}{10}}
\makeatother

\renewcommand{\thefootnote}{\fnsymbol{footnote}}
\renewcommand\footnoterule{\vspace*{1pt}%
\color{cream}\hrule width 3.5in height 0.4pt \color{black}\vspace*{5pt}} 
\setcounter{secnumdepth}{5}

\makeatletter 
\renewcommand\@biblabel[1]{#1}            
\renewcommand\@makefntext[1]%
{\noindent\makebox[0pt][r]{\@thefnmark\,}#1}
\makeatother 
\renewcommand{\figurename}{\small{Fig.}~}
\sectionfont{\sffamily\Large}
\subsectionfont{\normalsize}
\subsubsectionfont{\bf}
\setstretch{1.125} 
\setlength{\skip\footins}{0.8cm}
\setlength{\footnotesep}{0.25cm}
\setlength{\jot}{10pt}
\titlespacing*{\section}{0pt}{4pt}{4pt}
\titlespacing*{\subsection}{0pt}{15pt}{1pt}

\fancyfoot{}
\fancyfoot[RO]{\footnotesize{\sffamily{1--\pageref{LastPage} ~\textbar  \hspace{2pt}\thepage}}}
\fancyfoot[LE]{\footnotesize{\sffamily{\thepage~\textbar\hspace{3.45cm} 1--\pageref{LastPage}}}}
\fancyhead{}
\renewcommand{\headrulewidth}{0pt} 
\renewcommand{\footrulewidth}{0pt}
\setlength{\arrayrulewidth}{1pt}
\setlength{\columnsep}{6.5mm}
\setlength\bibsep{1pt}

\twocolumn[
  \begin{@twocolumnfalse}
\sffamily
\begin{tabular}{m{0cm} p{16.5cm} }

  & \noindent\LARGE{\textbf{Hydrodynamic Effects on the Motility of Crawling Eukaryotic Cells}} \\
\vspace{0.3cm} & \vspace{0.3cm} \\

 & \noindent\large{Melissa H. Mai \textit{$^{a}$} and Brian A. Camley\textit{$^{a b}$}} \\

& \noindent\normalsize{Eukaryotic cell motility is crucial during development, wound healing, the immune response, and cancer metastasis. Some eukaryotic cells can swim, but cells more commonly adhere to and crawl along the extracellular matrix. We study the relationship between hydrodynamics and adhesion that describe whether a cell is swimming, crawling, or combining these motions. Our simple model of a cell, based on the three-sphere swimmer, is capable of both swimming and crawling. As cell-matrix adhesion strength increases, the influence of hydrodynamics on migration diminish. Cells with significant adhesion can crawl with speeds much larger than their nonadherent, swimming counterparts. We predict that, while most eukaryotic cells are in the strong-adhesion limit, increasing environment viscosity or decreasing cell-matrix adhesion could lead to significant hydrodynamic effects even in crawling cells. Signatures of hydrodynamic effects include dependence of cell speed on the medium viscosity or the presence of a nearby substrate and the presence of interactions between noncontacting cells.  These signatures will be suppressed at large adhesion strengths, but even strongly adherent cells will generate relevant fluid flows that will advect nearby passive particles and swimmers.} \\

\end{tabular}

 \end{@twocolumnfalse} \vspace{0.6cm}

  ]

\renewcommand*\rmdefault{bch}\normalfont\upshape
\rmfamily
\section*{}
\vspace{-1cm}


\footnotetext{\textit{$^{a}$~Department of Biophysics, Johns Hopkins University, Baltimore, Maryland}}
\footnotetext{\textit{$^{b}$~Department of Physics and Astronomy, Johns Hopkins University, Baltimore, Maryland}}



\section*{Introduction}
Throughout development, wound healing, and cancer metastasis \cite{Birchmeier2003, Banchereau1998, Anon2012}, eukaryotic cells crawl while adherent to the extracellular matrix \cite{friedl2012new,friedl1998cell}. An increasing amount of evidence shows that eukaryotic cells without strong adhesion can also swim \cite{VanHaastert2011, Charras2014, Barry2010,franz2018fat,oneill2018membrane}; 
this is distinct from other mechanisms of adhesion-independent cell motion, e.g. ``chimneying'' \cite{Hawkins2009,lammermann2008rapid} or  osmotic engines \cite{stroka2014water,li2018transition}. In particular, Aoun and coworkers have observed lymphocytes directly transitioning between crawling on adhesive and swimming over non-adhesive regions of substrate, showing that cells may exploit both strategies depending on their environment \cite{Aoun2019}. We use a minimal model incorporating both hydrodynamics and regulated substrate adhesion to understand what happens when cells are intermediate between swimming and crawling.

At the micron length scales typical for eukaryotic cells, they must swim through fluids at low Reynolds number, where inertial forces become irrelevant. A low-Reynolds number swimmer is strongly constrained by the linearity and reversibility of the Stokes equations and can only achieve a net displacement if its motion is defined by a nonreciprocal, or time-irreversible, cycle of conformational changes \cite{Lauga2009a, Purcell1977}. As a result, a microswimmer needs at least two degrees of freedom to achieve a productive cycle of motion -- the so-called ``Scallop Theorem.''

\begin{figure}[t!]
	\centering
	\includegraphics[width=0.45\textwidth]{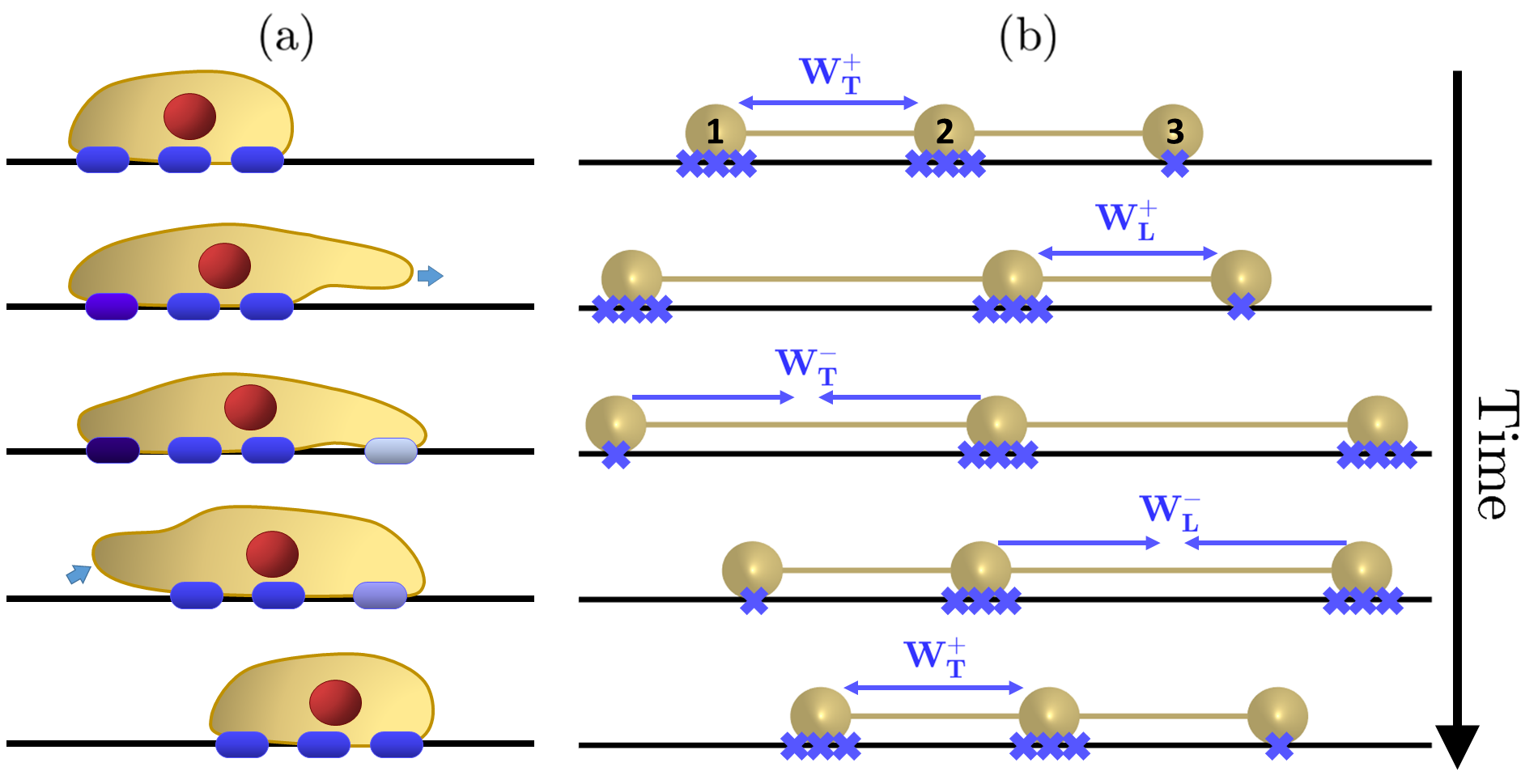}
	\caption{The motion sequence of our three sphere crawler (b) is chosen to resemble lamellipodial migration (a): the adhesion of each bead (blue hashes) depends on the current motion to model the maturation and rupture of adhesive contacts. The crawler's arms deform with prescribed velocities $W_L^{\pm}$ (leading arm) and $W_T^{\pm}$ (trailing arm). In an expansion phase, the cell arms expand from length $L-\Delta L/2$ to length $L+\Delta L/2$, and in contraction vice versa; the geometric parameters $L$ and $\Delta L$ and other parameters are listed in Table \ref{table:stdpar}.} \label{fig:3SphereModel}
\end{figure}

Eukaryotic cells crawl on substrates by cycles of extending protrusions at the cell front and contracting the cell body at the rear (\fig \ref{fig:3SphereModel}a). Forward protrusions attach through complexes of adhesion proteins, and contractions are aided by the motor protein myosin through rupturing the adhesive bonds in the rear \cite{Mitchison1996, Puklin-Faucher2009, Rafelski2004, Mogilner2002, Achard2010, Aranson2016, Fournier2010}. 
Since cell-substrate adhesion may be regulated to differ between the cell's back and front, crawling cells can violate the Scallop Theorem, allowing minimal models of cells as dimers to crawl \cite{Lopez2014,wagner2013crawling}. 

To capture swimming motion, our model must have at least two degrees of freedom. We adapt the classical three-sphere swimmer \cite{Najafi2004, Golestanian2008}, describing our cell as three beads connected by two arms. These arms extend and contract around a mean arm length $L$ in a nonreciprocal sequence with prescribed distortion velocities (\fig \ref{fig:3SphereModel}b). 
Significant work has been done to characterize theoretical three-sphere swimmers \cite{Golestanian2008b, Taghiloo2013, Box2017}, including their interactions with walls \cite{Zargar2009, Daddi-Moussa-Ider2018, Or2011} and swimmer-swimmer interactions \cite{Pooley2007, Farzin2012}. Three-sphere swimmers have even been built experimentally with optical tweezers \cite{Leoni2009}. 

Here our approach is to use a minimal model, neglecting many details of biochemistry and cell shape which have been studied extensively for crawling cells \cite{Aranson2016,Tjhung2015,keren2008mechanism,Shao2012,Camley2017,Ziebert2013,albert2016dynamics,holmes2012comparison} and more recently also for swimming ones \cite{wu2016amoeboid,Aoun2019,campbell2017computational} (and in a very recent example, a transition between confined crawling and swimming \cite{noselli2019swimming}). In our three-sphere crawler, we describe adhesion to the surface by introducing an adhesive drag force, and we examine the relationships between adhesion and the hydrodynamics of swimming and crawling cells. Finally, we examine the hydrodynamic interactions among multiple crawlers and swimmers. 

\section*{Model and Methods}

We describe our cell with the minimal structure of three beads, representing the tail, body, and head of the cell (labeled 1, 2, and 3 in \fig \ref{fig:3SphereModel}). We prescribe the relative motion of the head and tail of the cell to match the stereotypical cycle of protrusion and retraction, as shown in \fig \ref{fig:3SphereModel}. This differs from earlier models of cell crawling that generally prescribe the forces driving cell motion, then find the resulting cell velocities \cite{Ziebert2013,Lopez2014,Camley2017}. Instead, we set the velocities of the cell head and tail relative to the body and then solve for the forces that obey physical constraints of zero net internal force and torque (see below); this is a more typical approach for modeling swimming \cite{Lauga2009}, and allows our model to limit back to the classical three-sphere swimmer at zero adhesion.

We also include friction-like adhesion forces between the cell and substrate, the strength of which will control whether the cell's motion is primarily driven by swimming or crawling. Cell-substrate adhesion is tightly regulated, and so we choose these adhesion strengths to depend on the cycle of the motion (Table \ref{table:xivals}), allowing the cell to crawl. These motions are also chosen to be nonreciprocal \cite{Lauga2009,Purcell1977}, so that in the absence of adhesive force, the cell may still swim. Below, we describe how we solve the Stokes equations that describe fluid flow, how we model cell-substrate adhesion, the physical constraints on the cell, and the time-stepping algorithm we use to evolve the cell's motion. 

\subsection*{Hydrodynamic model of forces and motion}\label{sec:overview}
We describe cell motion in a fluid environment by relating the forces applied to the model's three beads to their velocities. Cells move in a low Reynolds number environment where viscous drag forces dominate fluid motion and inertial forces become irrelevant \cite{Lauga2009,Purcell1977}. In this regime, the fluid flow surrounding a motile cell is described by the time-independent Stokes equations for incompressible fluids, which describe the velocity of a point at $\vb{r}$ in a fluid with pressure $p$ subject to force density $\vb{f}(\vb{r})$ \cite{Kim2016}:
\begin{align}
\eta\nabla^2\vb{v}(\vb{r}) &= \nabla p(\vb{r}) - \vb{f}(\vb{r})\label{eq:Stokes1}\\
\nabla \cdot \vb{v}(\vb{r}) &= 0\label{eq:Stokes2}
\end{align}

If the force density is a tightly-localized point, with $\vb{f}(\vb{r}) = \vb{F} \delta(\vb{r})$, the Stokes equation can be solved as
\begin{equation}
\vb{v}(\vb{r}) = \ten{G}(\vb{r})\cdot \vb{F}
\end{equation}
where $\ten{G}(\vb{r})$ is the Green's function of the Stokes equations, known as the Oseen tensor. In components, this equation is $v_\alpha(\rb) = G_{\alpha\beta}(\rb) F_\beta$.  We assume Einstein summation here and throughout the paper. In an unbounded, three-dimensional fluid,
\begin{equation}
G_{\alpha\beta}(\vb{r}) = \frac{1}{8\pi\eta}\left(\frac{\delta_{\alpha\beta}}{r} + \frac{r_\alpha r_\beta}{r^3}\right) \; \; \;\textrm{(unbounded fluid)} \label{eq:oseen}
\end{equation}
where $\alpha,\beta = x,y,z$ are the Cartesian coordinates. Generalizations of this Oseen tensor can be made to different boundary conditions \cite{Blake1971,Kim2016}.  \eq \ref{eq:oseen} diverges as $r\to0$, suggesting the velocity of a point subject to a point force is ill-defined. We handle this through the method of regularized Stokeslets \cite{Cortez2001,Cortez2005}, smearing the point force over a scale $\epsilon$. In this approach, we assume that the force distribution over a bead is $f(\rb) = \vb{F} \phi_\epsilon(\rb)$, where $\phi_\epsilon(\rb)$ is a radially-symmetric ``blob'' that integrates to one. In this case, 
\begin{equation}
\vb{v}(\vb{r}) = \ten{G}(\vb{r};\epsilon)\cdot \vb{F}
\end{equation}
where now the {\it regularized} response $G_{ij}(\rb;\epsilon)$ remains finite as $\rb \to 0$. $G_{ij}(\rb;\epsilon)$ depends on the choice of $\phi_\epsilon(\rb)$; several variants are discussed in \cite{Cortez2001,Cortez2005,Ainley2008,leiderman2016swimming,hernandez2007fast,camley2013diffusion,noruzifar2014calculating}. 

By the linearity of the Stokes equations, the velocity in response to many regularized forces is a superposition of these solutions
\begin{equation}
\vb{v}(\vb{r}) = \sum_n \ten{G}(\vb{r}-\vb{R_n};\epsilon)\cdot \vb{F}(\vb{R}_n) \label{eq:flow}
\end{equation}
If the forces are known, the velocity of bead $m$ is
\begin{equation}
\vb{v}(\vb{R}_m) = \sum_n \ten{G}(\vb{R}_m-\vb{R}_n;\epsilon)\cdot \vb{F}(\vb{R}_n) \label{eq:beadbead}
\end{equation}
If there are $n_b$ blobs in our system (composed of one or many cells, with three blobs per cell), \eq \ref{eq:beadbead} can be thought of as a set of $3 n_b$ linear equations giving the bead velocity components in terms of the force components,
\begin{equation}
\vb{V} = \hat{M}\vb{F}\label{eq:VMFsimple}
\end{equation}
where $\hat{M}$ is a $3n_b \times 3 n_b$ mobility matrix defining the hydrodynamic interactions among the spheres. 

We will treat two hydrodynamic geometries in this paper: 1) cells with no hydrodynamic obstruction, and 2) cells near a solid surface, when we will use the regularized Stokeslet solution of Ainley \textit{et al.}, which creates the response to a point force near a no-slip wall from a superposition of higher-order solutions to the Stokes equations, defined in the Supplemental Material \cite{Ainley2008}.  This is a regularization of the solutions by Blake \cite{Blake1971}, which has been previously used to study the behavior of swimmers near walls \cite{Zargar2009, Daddi-Moussa-Ider2018,Or2011, Simha2018, Spagnolie2012}. {For cells away from a solid substrate (Case 1), we simply take the regularized Stokeslet of \cite{Ainley2008} in the limit of cells far from the wall.}

\begin{table*}[tb]
	\caption{The adhesive friction, $\xi_i$, for the trailing (1), center (2), and leading (3) beads are qualitatively described for each motion. In the simulations used for this work, $\xi_{high} = \xi$ and $\xi_{low} = 0.2\xi$, where $\xi$ is the global adhesion parameter. Finally, the deformation velocities for the leading ($W_L$) and trailing ($W_T$) arms are, for simplicity, chosen to be $\pm W$ or $0$ for each motion.}
	\centering
	\begin{tabular*}{0.8\textwidth}{c @{\extracolsep{\fill}} c c c c c}
		\hline
		\textit{Motion} & $\xi_{1}$ & $\xi_{2}$ & $\xi_{3}$ & $W_L$ & $W_T$\\
		\hline
		Trailing arm extension & High & High & Low & $0$ & $+W$\\
		Leading arm extension & High & High & Low & $+W$ & $0$\\
		Trailing arm contraction & Low & High & High & $0$  & $-W$\\
		Leading arm contraction & Low & High & High & $-W$ & $0$\\
		\hline
	\end{tabular*}
	\label{table:xivals}
\end{table*}
\subsection*{Adhesion forces}
\label{sec:adhesion}
To model the effect of protein-mediated cell adhesion to a substrate or fiber, we introduce an adhesive force, $\Fadh$, by the substrate on each bead in the form of a frictional drag:
\begin{equation}
\Fadh(t) = -\boldsymbol{\xi}(t)\circ\vb{V}
\end{equation}
where $\boldsymbol{\xi}(t)$ is a $3 n_b\times 1$ column vector defining the adhesive friction coefficients for each bead in each direction depending on the current motion (and therefore time: see Table \ref{table:xivals}).   The symbol $\circ$ represents Hadamard (elementwise) multiplication: the components of the force are $F_i^{\textrm{adh}} = -\xi_i(t) V_i$. Here $i$ is a generalized index going over both bead and dimension, i.e. $i = 1x, 1y, 1z, 2x\cdots$. This linear form is appropriate in the low-speed limit of motion over the substrate \cite{srinivasan2009binding,sabass2010modeling,li2010model}. We choose the drag force to be tangential to the substrate, i.e. the $x$ and $y$ components for each bead are equal to each other, but the $z$ component is set to zero. This is irrelevant in practice, since we will assume that an adherent cell is constrained to not move in the $z$ direction. 

Because this frictional drag force is linear in the velocity, we can derive a simple form for the velocity even in the presence of this additional drag. Assuming that the {\it total} force in \eq \ref{eq:VMFsimple} is composed of cell-internal forces $\Fint$ and the cell-substrate friction, i.e. $\vb{F} = \Fint + \Fadh(t)$, we find
\begin{align*}
\vb{V} &= \hat{M}\left(\Fint + \Fadh(t)\right) \\
&= \hat{M}\Fint - \hat{M}\left(\boldsymbol{\xi}(t)\circ\vb{V}\right)
\end{align*}
which implies
\begin{equation}
\mathbb{I}\vb{V} +  \hat{M}\left(\boldsymbol{\xi}(t)\circ\vb{V}\right) = \hat{M}\Fint\numberthis\label{eq:VMFfull}
\end{equation}
where $\mathbb{I}$ is the identity matrix. The term $\hat{M}\left(\boldsymbol{\xi}(t)\circ\vb{V}\right)$ is just a matrix multiplying $\vb{V}$:
\begin{equation}
\left[\hat{M}\left(\boldsymbol{\xi}(t)\circ\vb{V}\right)\right]_i = M_{ij} (\xi_j V_j) \equiv \left[\hat{\Xi}\vb{V}\right]_i  \label{eq:XiMat}
\end{equation}
where
\begin{equation}
\Xi_{ij} = M_{ij}\xi_j \label{eq:XiEinstein}
\end{equation}
As a result, Equation (\ref{eq:VMFfull}) becomes
\begin{equation}
\vb{V} = \left(\mathbb{I} + \hat{\Xi}\right)^{-1}\hat{M}\Fint\label{eq:VMFXi} 
\end{equation}
We define the \textit{modified mobility matrix,} $\hat{\mathcal{M}}$, such that
\begin{align}
\hat{\mathcal{M}} &= \left(\mathbb{I} + \hat{\Xi}\right)^{-1}\hat{M} \label{eq:modifiedM}\\
\vb{V} &= \hat{\mathcal{M}}\Fint. \label{eq:VMFfinal}
\end{align}
The value of \eq \ref{eq:VMFfinal} is that we can now directly relate the velocities and the internal forces, without needing to handle the adhesion forces explicitly. This is useful because some of our physical constraints apply only to the internal forces -- such as the requirement that each cell cannot exert a net internal force on itself.

Depending on the phase of the cell's motion (Table \ref{table:xivals}), we choose the components of $\boldsymbol{\xi}(t)$ to be either $\xi^\textrm{high} = \xi$ or $\xi^\textrm{low} = 0.2 \xi$, where $\xi$ is the overall scale of the adhesion. For example, during leading edge extension, adhesion at the front is low since the focal contacts have not yet matured, yet the adhesion in the rear is strong. During trailing edge contraction, the rupture of focal contacts and targeted disassembly \cite{broussard2008asymmetric} causes the adhesion on the rear bead to be lower, while other parts of the cell are more strongly bound to the surface. We only model the switching between ``high'' and ``low'' adhesion strengths during the cycle -- intermediate values can also be used but require further parametrization and produce qualitatively similar results.

\subsection*{Constraints}\label{sec:constraints}

Defining a cell's motion via \eq \ref{eq:VMFfinal} requires knowledge of all the internal forces, which can be found by applying the necessary constraints on the cell's motion. We have three important sets of constraints: 1) the pattern of extension of the cell front and back, 2) the linear geometry of the three-bead cell, where we apply the appoach of \cite{Daddi-Moussa-Ider2018}, and 3) no unphysical forces or torques required on the cell. We enforce the linearity constraint as a minimal model for the cell's internal resistance to deformation. These constraints will be different for adherent cells (those attached to a surface) and non-adherent cells (those just swimming near a surface).

\subsubsection*{Constraints for non-adherent cells}\label{sec:swimcst}
For non-adherent cells ($\xi = 0$), $\vb{F} = \Fint$, and our model is just a three-bead swimmer, as in e.g. \cite{Daddi-Moussa-Ider2018}. For a fully and uniquely determined system, defining the nine components of $\vb{F}$ per cell (three for each bead) requires nine constraints.

We require that the cell is not generating a net internal force (``force-free'' \cite{Lauga2009}), providing three independent constraints:
\begin{equation}
\sum_{n=1}^{3}\vb{F}_n^\textrm{int} = \boldsymbol{0}\label{eq:forcefree}
\end{equation}

The cell also cannot generate a net internal torque:
\begin{equation}
\sum_{n = 1,2,3} (\vb{R}_n - \vb{R}_2) \times \vb{F}_n^\textrm{int} = \boldsymbol{0}\label{eq:torquefree}
\end{equation}
where, for convenience, the reference point for calculating the torques is set as the center sphere's position. Due to the symmetry of this system and the rigid-body constraints discussed later, the component of the torque along the cell's axis is always zero, so only two components of the torque constraint are independent and enforced. When simulating multiple cells at a time, each cell is required to be individually force- and torque-free. 

\begin{figure}[t!]
	\centering\includegraphics[width=0.25\textwidth]{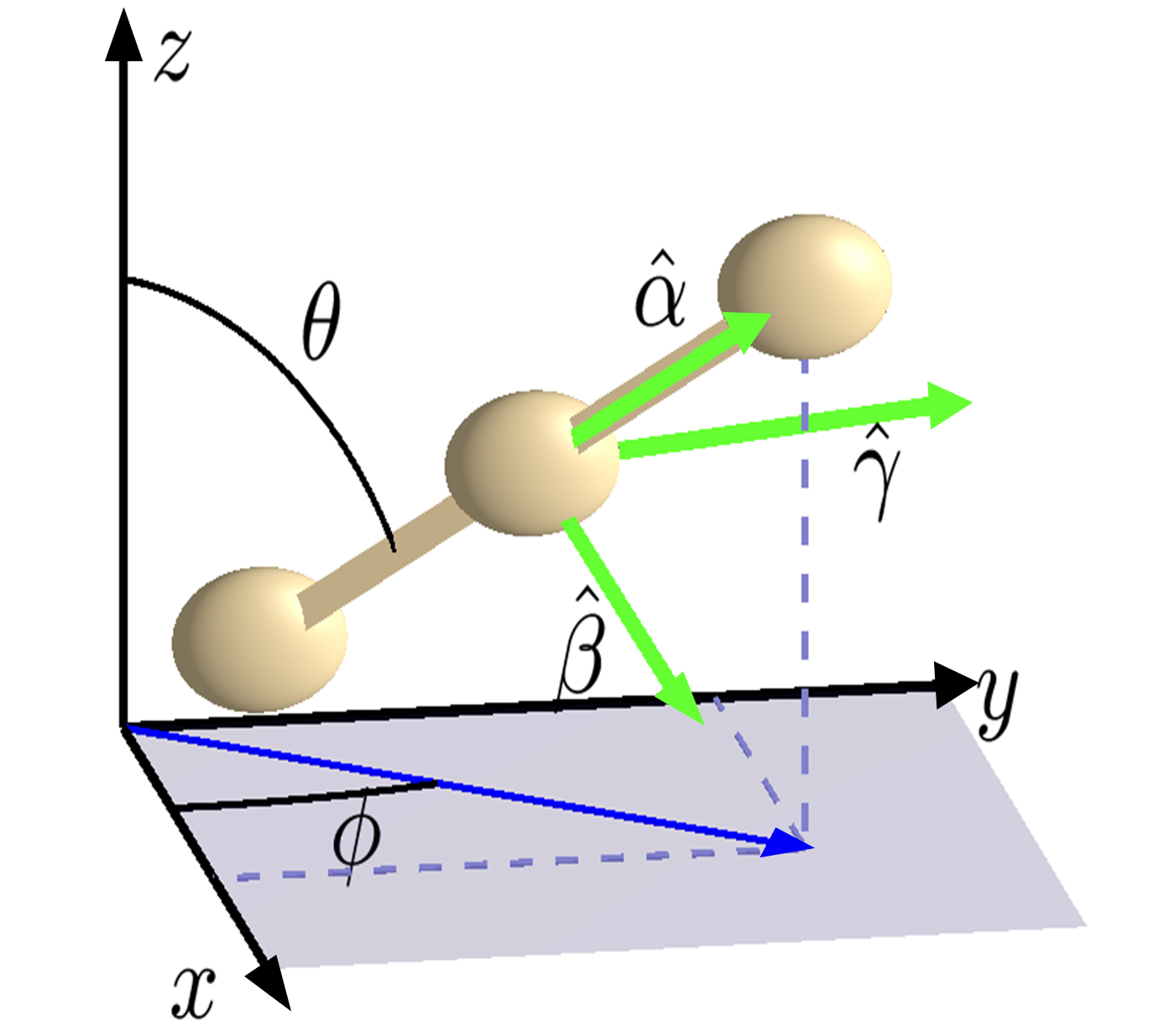}
	\caption{The cell's orientation defines an orthonormal basis $\lbrace\hat{\alpha},\hat{\beta},\hat{\gamma}\rbrace$ (green) depending on the angles $\theta$ and $\phi$.
	}\label{fig:orientation}
\end{figure}

To keep track of the cell's orientation, we define a rotated set of orthonormal basis vectors as in \cite{Daddi-Moussa-Ider2018} (\fig\ref{fig:orientation}): 
\begin{align}
\hat{\alpha} &= (\sin\theta\cos\phi, \sin\theta\sin\phi, \cos\theta)^T\label{eq:that}\\
\hat{\beta} &= (\cos\theta\cos\phi, \cos\theta\sin\phi, -\sin\theta)^T\label{eq:thetahat}\\
\hat{\gamma} &= (-\sin\phi, \cos\phi, 0)^T\label{eq:phihat}
\end{align}
where $\theta$ and $\phi$ are the polar and azimuthal angles, respectively. Here, $\hat{\alpha}$ is the cell migration direction, and $\hat{\beta}$ and $\hat{\gamma}$ are two convenient vectors normal to the cell's direction. Hence in this basis, as discussed above, only the projections of the torque onto $\hat{\beta}$ and $\hat{\gamma}$ are explicitly constrained. 

The final four constraints arise from fixing the deformation of the arms and the rigid body constraint. In defining the motion of the cell, we choose the deformation velocity of the leading arm (connecting bead 3 and 2) to be $W_L$ and that of the trailing (connecting 2 and 1) to be $W_T$, which both depend on the phase of motion. The motion is along the axis of the crawler, so projections of the relative velocities onto the principal orientation vector $\hat{\alpha}$ should be equal to the deformation velocities:
\begin{align}
(\vb{V}_3 - \vb{V}_2)\cdot\hat{\alpha} &= W_L \label{cst:WL}\\
(\vb{V}_2 - \vb{V}_1)\cdot\hat{\alpha} &= W_T \label{cst:WT}
\end{align}
The projections onto the other two orientation vectors enforce the rigid body constraint of no internal bending -- the projections of the change in the orientation and length of both arms onto $\hat{\beta}$ should be equal and opposite. Specifically,
\begin{align}
L_L^{-1}(\vb{V}_3 - \vb{V}_2) \cdot \hat{\beta} &= -L_T^{-1}(\vb{V}_1 - \vb{V}_2)\cdot\hat{\beta}\label{cst:beta}
\end{align}
The same constraint is applied to the projection onto $\hat{\gamma}$:
\begin{align}
L_L^{-1}(\vb{V}_3 - \vb{V}_2) \cdot \hat{\gamma} &= -L_T^{-1}(\vb{V}_1 - \vb{V}_2)\cdot\hat{\gamma}\label{cst:gamma}
\end{align}
These constraints are linear equations for the velocity, even though our earlier constraints are linear equations for the forces $\vb{F}$. We can convert these to linear equations for $\vb{F}$ via \eq\ref{eq:VMFfinal}. (See Supplemental Material for detailed explanation). 

\subsubsection*{Adherent cells}\label{sec:crawlcst}
We assume that an adherent cell ($\xi > 0$) does not move away from the substrate -- it is strongly attached. Instead of explicitly modeling an attachment force, we handle this by constraining the $z$-directional velocity for each bead to be zero:
\begin{equation}
\vb{V}_n \cdot \hat{\vb{z}} = 0 \; \; \; n = 1,2,3
\end{equation}
A crawler will thus stay at a fixed distance away from the substrate ($z = 0$); we choose the crawler to be at height $z = a$, i.e. with the spheres resting on the surface. Again, this constraint on the velocity can be converted into a constraint on the forces through \eq \ref{eq:VMFfinal}. Including these three additional constraints requires relaxation of three constraints from the non-adherent case. This avoids mathematical overdetermination of the system and physical redundancy of constraints for an adherent cell. 

We relax the $z$-component of the force-free condition, as there must be some vertical force keeping the cell bound to the surface. Additionally, we remove the now-redundant constraint of  \eq \ref{cst:beta}, since attachment to the surface mandates fixing of the polar angle to $\theta = \pi/2$. Finally, we remove the projection of the torque-free condition onto $\hat{\gamma}$ due to the relaxation of the force-free condition's $z$-component.

\subsubsection*{The constraint matrix}\label{sec:cstmtx}
Once all of the constraint equations have been written in terms of the forces $\Fint$, we will have an equation of the form
\begin{equation}
\hat{C}\Fint=\vb{d}\label{eq:Cmat}
\end{equation}
where $\hat{C}$ and $\vb{d}$ for both adherent and non-adherent cases are explicitly defined in the Supplemental Material. We solve \eq \ref{eq:Cmat} using LU factorization (MATLAB's $\texttt{linsolve}$).

\begin{figure*}[b!]
	\centering
	\includegraphics[width=0.7\textwidth]{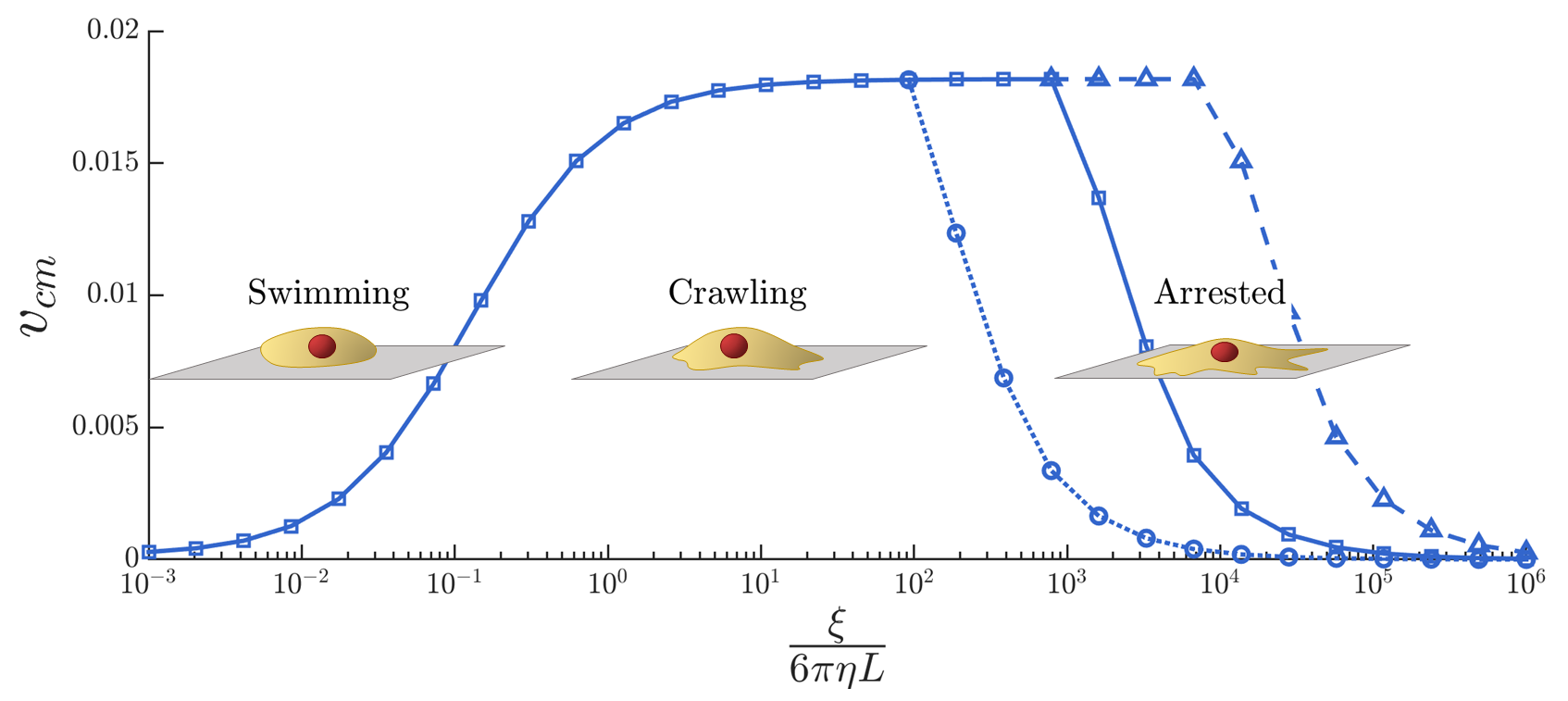}
	\caption{A typical velocity profile for a three-sphere cell over $\xi/6\pi\eta L$ exhibits a biphasic dependence on adhesion strength. At low adhesion, the cell exhibits a slipping, swimming behavior. As adhesion increases, the cell is better able to grip the surface and crawl, until the adhesion becomes too strong, which leads to arrested migration. Decreasing (dotted, circles) and raising (dashed, triangles) the threshold force $\Fthresh$ moves the turning point linearly in the appropriate direction. Profiles were generated for $\Fthresh = 10^2$, $10^3$, and $10^4$. Other parameters are as in Table \ref{table:stdpar}.}\label{fig:VCurveStd}
\end{figure*}
\subsection*{Threshold force}\label{sec:fthresh}
In our simulation, we prescribe the relative motion of the front and back of the cell and solve for the forces needed to move at this speed. However, the internal force required increases with increasing adhesion, yet a cell can only exert a finite amount of force. We apply a simple limitation on the internal force exerted on any bead. If the prescribed deformation of the arms requires an internal force $\Freq$ of a magnitude greater than a threshold force $\Fthresh$, the cell exerts only its maximal force, resulting in a linear scaling of {\it all} internal forces:
\begin{equation}
\Fint = \Freq\cdot\frac{\Fthresh}{\max |\Freq|}\label{eq:Fscale}
\end{equation}
This constraint is enforced after defining the matrix in \eq \ref{eq:Cmat}. Since the only nonzero terms of $\vb{d}$ refer to the deformation velocity constraints (\eq \ref{cst:WL} and \ref{cst:WT}), any linear scaling of $\Fint$ still satisfies all other constraints. As a result, the scaled forces continue to obey all the necessary physics of the system but reduce the cell's overall center-of-mass velocity.

If the determined velocities are scaled down in this manner, we adapt the time step used as 
$\Delta t' = \Delta t \cdot \frac{\max |\Freq|}{\Fthresh}$.
This allows the same deformation length to occur during each iteration, reducing the computational cost of the simulation.

\subsection*{Algorithm}\label{sec:algorithm}
We employ a time-stepping algorithm to numerically solve the problem, outlined below. The parameters used for the simulations are presented in the Supplemental Material. The mobility tensors, forces, and velocities are reevaluated at each step.
\begin{enumerate}
	\item Determine which arm is extending and/or contracting, and define the appropriate adhesion strength (Table \ref{table:xivals})
	\item Calculate the modified mobility matrix, $\hat{\mathcal{M}}$
	\item Construct the constraint matrix, $\hat{C}$
	\item Find required forces (\eq \ref{eq:Cmat}), scale by $\Fthresh$ if needed
	\item Calculate velocities $\vb{V}$ via \eq\ref{eq:VMFfinal}
	\item Update the configuration via Euler's method with a defined time step $\Delta t$: $\vb{R}(t + \Delta t) = \vb{R}(t) + \vb{V}\Delta t$
\end{enumerate}

\subsection*{Parameter setting} 
Throughout this paper, we will use convenient units of mean cell arm length $L = 1$, arm speeds $W = 0.1$, and fluid viscosity $\eta = 1$. To map between our simulation units and experimental measurements for different cells, we must have estimates for these different numbers, as well as for the friction coefficient $\xi$ and the threshold force $\Fthresh$. Fibroblasts on nanofibers have a protrusive velocity of order $0.1\mu$m/s \cite{Guetta-Terrier2015}, so in this context, our units of velocity can be interpreted as $\mu$m/s; the maximum velocities of order $0.02$ in simulation units correspond to speeds of $\sim 70 \mu$m/hr, consistent with \cite{Guetta-Terrier2015}. 

If we assume $\eta = 10^{-3}$ Pa s is the viscosity of water and $L = 100 \mu$m (order of magnitude correct for fibroblasts \cite{Guetta-Terrier2015,Doyle2009}, though they can be very long in narrow confinement or on fibers), our simulation unit of force corresponds to $10^{-3} \, \textrm{Pa s} \times 1 \mu\textrm{m s}^{-1} \times 100 \mu\textrm{m} = 0.1 \textrm{pN}$. We expect these maximum forces to be on the order of nanonewtons, so this suggests $\Fthresh \approx 10^3 - 10^5$ in simulation units. Similarly, one simulation unit of the drag $\xi$ is $10^{-3} \textrm{Pa s} \times 100 \mu \textrm{m} = 0.1 \mu \textrm{m Pa s} = 10^{-4} \textrm{nN}/(\mu \textrm{m/s})$. 

The only remaining variable to be set is the drag coefficient $\xi$: this is a difficult parameter to estimate, and in general we will vary $\xi$ over a broad range and see what consequences follow. We make an initial, rough estimate by using data from traction stress experiments on keratocytes. Ref. \cite{Fournier2010} found a linear relationship between actin velocity $v$ and substrate stress $\sigma$ of the form $\sigma = k v + \sigma_0$, with $k \sim 0.2-1 \textrm{kPa} / (\mu \textrm{m/s})$. {We estimate $\xi$ as the product of $k$ with the contact area of one section of the cell, $A \approx 20 \mu m^2$, or $\xi \sim 10 \textrm{nN} / (\mu m/s)$. This suggests that in our simulation units, strongly adherent cells will have a friction coefficient of $\xi \approx 10^5$. }

When we are below the threshold force, the dynamics of our crawler will be largely controlled by the relative importance of hydrodynamic flow and adhesion. We characterize this with the unitless parameter $\xi/6 \pi \eta L$. {Strongly adherent cells will have $\xi/6\pi\eta L \approx 5000.$ Cells with weaker adhesion (e.g. \textit{Dictyostelium} amoebae or cells on less adhesive substrates) or cells in more viscous environments will have a stronger relative importance of hydrodynamics.}

The specific parameters used in each figure are presented either in the figure or in Tables \ref{table:stdpar}-\ref{table:kicker}.

\section*{Results}

\begin{figure}[t!]
	\centering
	\includegraphics[width=0.45\textwidth]{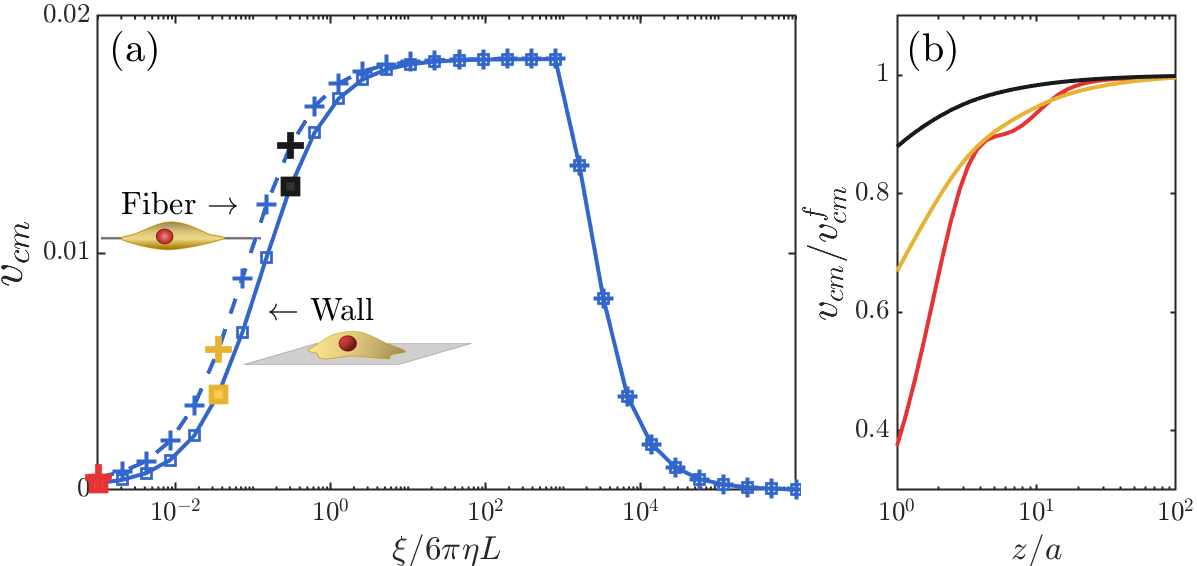}
	\caption{Substrate hydrodynamic effects. (a) At low adhesion, wall-induced hydrodynamic drag (solid, $z/a = 1$) slows a cell relative to its motion on or near a thin fiber (dashed, $z/a > 100$), but high-adhesion motion is unaffected by substrate hydrodynamics. (b) Center-of-mass velocity, scaled by the uninhibited (fiber) velocity, at different adhesive strengths corresponding to the colors in (a), as a function of cell height above the wall $z$, scaled by the bead radius $a$. Flow fields for cells at different distances above the substrate are shown in \fig \ref{fig:wall_flows}. Parameters are listed in Table \ref{table:stdpar}.
	}\label{fig:Substrate}
\end{figure}
\subsection*{Biphasic dependence of migration speed on adhesion strength}\label{sec:res:orig}
A typical velocity profile of a migrating three-sphere cell is shown in \fig \ref{fig:VCurveStd}. The model captures the biphasic dependence on adhesion strength that has been observed experimentally \cite{Barnhart2011,gupton2006spatiotemporal}. A weakly-adherent cell slips along the surface, essentially swimming. Movement in this limit is slow -- low Reynolds number swimming is typically quite inefficient \cite{Purcell1977}. As adhesion increases, the cell can better grip the surface and drag itself along, with velocity increasing until a plateau at a value roughly sixty times greater than the swimming speed. 
At sufficiently high adhesion, to maintain its motion, the cell would have to exceed the threshold force $\Fthresh$. Constrained by this maximal force, the cell must slow down and eventually stop moving. 

The position of the stalling transition at high adhesion can be modulated via changing \Fthresh. In the case of a single crawler where the forces are dominated by adhesion, we can exactly solve for the motion using an approach similar to that of \cite{Golestanian2008}. We find, for the modulation of adhesion strengths outlined in Table \ref{table:xivals}, that the center-of-mass velocity is given by
\begin{equation}
\arraycolsep=1.4pt\def\arraystretch{2.0}
v_\textrm{cm} = W \frac{1-\alpha}{4+2\alpha}\left\{
\begin{array}{cc}
1 & \Fstar < \Fthresh \\
\frac{2\Fthresh}{\Fthresh+\Fstar}&  \frac{2\alpha}{1+\alpha}\Fstar \leq \Fthresh \leq \Fstar \\
2 \frac{1+\alpha}{1+3\alpha} \frac{\Fthresh}{\Fstar} & \Fthresh < \frac{2\alpha}{1+\alpha}\Fstar
\end{array}\right. \label{eq:analytic}
\end{equation}
where $\alpha = \xi_\textrm{low}/\xi_\textrm{high}$ ($\alpha = 0.2$ in our simulations), and $\Fstar = W\xi_\textrm{high}\frac{(1+\alpha)}{2+\alpha}$ is a characteristic force. (Details of derivation are in the Supplemental Material.) This result neglects all hydrodynamic interactions, but successfully describes the plateau in $v_{\textrm{cm}}$ and subsequent arrest.

We can see from \eq \ref{eq:analytic} that the plateau velocity $W \frac{1-\alpha}{4+2\alpha}$ depends only on the speed of protrusion $W$ and the ratio between the high and low levels of adhesion. Unsurprisingly, when there is no difference between the adhesion at the front and the back of the cell ($\alpha = \xi_\textrm{low}/\xi_\textrm{high} = 1$), the cell cannot crawl via adhesion. We can also identify the critical adhesion strength at which the cell begins to stall, the point at which $\Fstar = \Fthresh$, or $\xithresh =  \frac{2+\alpha}{1+\alpha} \Fthresh/W$.

\begin{figure}[t!]
	\centering
	\includegraphics[width=0.45\textwidth]{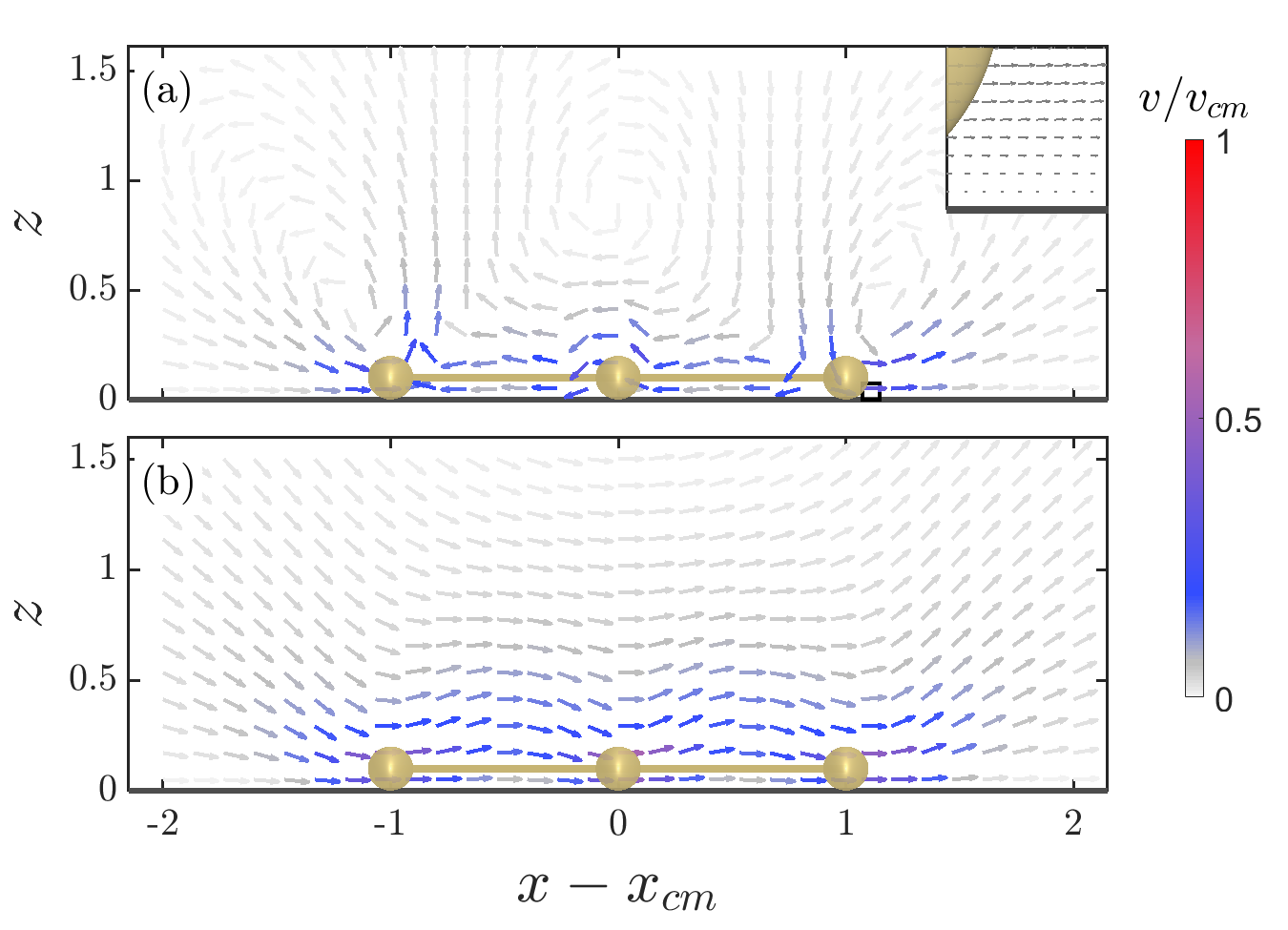}
	\caption{Time-averaged flow fields for a swimmer, $\xi/6\pi\eta L = 0$ (a) and a crawler, $\xi/6\pi\eta L = 10^3$ (b) with respect to the cell's center of mass on a wall located at $z = 0$. The color of the arrows corresponds to the magnitude of the velocity with respect to the cell's average center-of-mass velocity. The field shown is the flow field in the $xz$-plane, through the cell's axis. The inset shows that the fluid velocity vanishes near the wall to obey no-slip boundary conditions. Parameters are listed in Table \ref{table:stdpar}.}\label{fig:flowfield}
\end{figure}

\subsection*{Substrate hydrodynamics}\label{sec:res:substrhydro}
Because fluid cannot penetrate the substrate or slip past it, the substrate alters the hydrodynamic flow near the cell, altering swimming patterns, with attraction or repulsion depending on orientation and distance from the wall \cite{Berke2008, Guell1988,Dreyfus2005,Drescher2009}. Does the presence of a wall alter crawling speeds? We calculated velocity profiles for a cell crawling on a planar substrate or on an isolated fiber \cite{Guetta-Terrier2015,sheets2013shape,sharma2013mechanistic} (\fig\ref{fig:Substrate}a). We describe the fiber as infinitely thin with negligible hydrodynamic effects -- this would correspond to the limit of being infinitely far away from a supporting wall. We see that cells on substrates crawl more slowly than those on fibers -- but only when $\xi$ is sufficiently small. At large $\xi$, these hydrodynamic drag distinctions are negligible. 

This substrate-induced drag would also be present for a cell crawling along a fiber suspended above a substrate, as in the experiments of \cite{sharma2013mechanistic}, and might provide an experimental signature of hydrodynamics-dependent motility. We show that the speed of a cell on a fiber depends on distance from the substrate (\fig \ref{fig:Substrate}b). The influence of the wall depends on adhesion strength and the distance from the wall, vanishing almost entirely when the cell is 10$a$ above the surface, and with this distance becoming smaller at higher adhesions. Streamlines for crawlers on fibers at differing heights from the substrate are shown in \fig \ref{fig:wall_flows}. 

\subsection*{Crawlers generate fluid flow}\label{sec:res:flowfield}
While we have shown so far that hydrodynamic effects do not determine the cell's migration speed in the high adhesion limit, this does not mean that a crawling cell does not interact with its surrounding fluid. We calculate the flow field $\vb{v}(\rb)$ around a cell, averaged over five full motion cycles. We measure this flow field as a function of distance from the the cell's center of mass, finding $\vb{v}$ on a grid of points defined around the cell's center of mass using \eq \ref{eq:flow}. The time-averaged flow fields for a nonadherent cell ($\xi = 0$) and a strongly adherent crawler ($\xi/6\pi\eta L = 10^3$) on a wall are shown in \fig \ref{fig:flowfield}. 

While the swimmer produces a velocity field similar to a force quadrupole, the crawler behaves as three individual Stokeslets in the near-field limit and as a single Stokeslet far away; {this is particularly apparent when we simulate crawlers on fibers far from substrates (Figs. \ref{fig:farfield}-\ref{fig:farfield3})}. The critical difference between the crawler and the swimmer is that, because the crawler can exert force on a substrate, it can create a force monopole on the surrounding fluid without a net internal force, creating longer-range responses in flow than the three-sphere swimmer  \cite{Pooley2007}. 

\begin{figure*}[t!]
	\centering\includegraphics[width=0.9\textwidth]{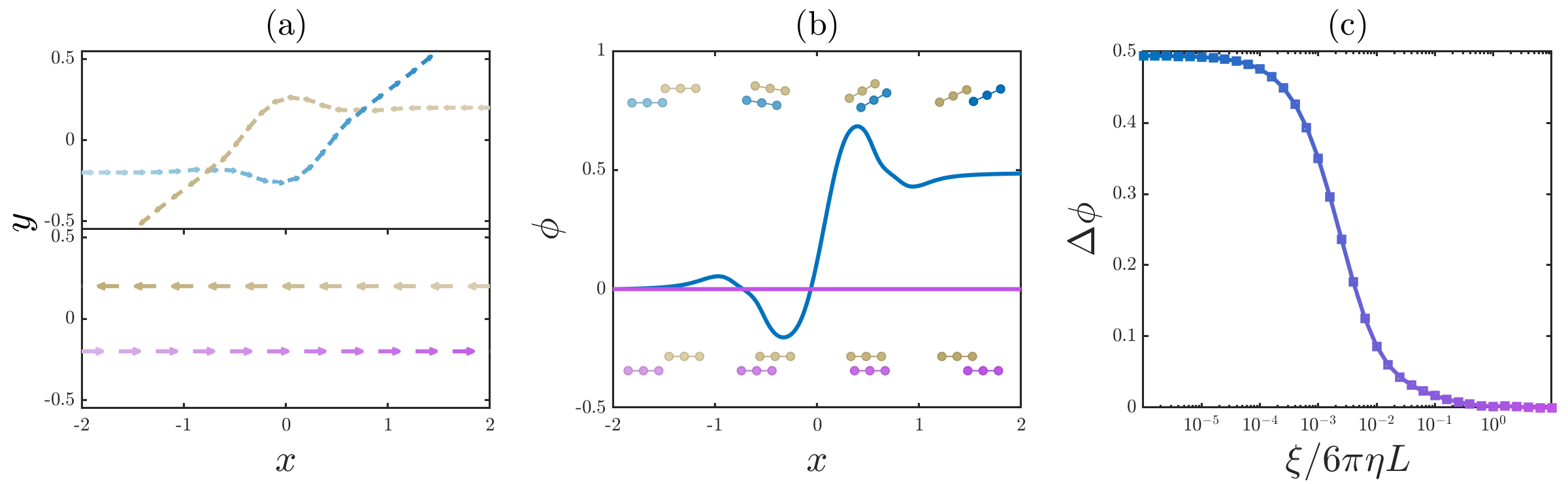}
	\caption{Hydrodynamic interactions of in-phase, antiparallel migrating cells that are adherent to a substrate (at height $z = a$). (a) Trajectories of two antiparallel cells at low (blue, $\xi/6\pi\eta L = 10^{-6}$) and high (purple, $\xi/6\pi\eta L = 10$) adhesion. (b) The azimuthal angle of the $+x$-oriented cell as the two cells pass one another. A schematic of the two cells are depicted at the corresponding positions in the trajectory. (c) Net deflection of the $+x$-oriented cell as a function of adhesion shows loss of hydrodynamic interactions at the high-adhesion limit. Parameters are listed in Table \ref{table:antipar}. }\label{fig:multicell}
\end{figure*}
\subsection*{Interactions between adherent cells}\label{sec:res:multicellall}

To quantify hydrodynamic interactions between cells, we study the trajectories of cells crawling toward each other on initially antiparallel paths separated by a distance of $0.4L$. We initially set the cell protrusion cycles to be in phase. In the low adhesion limit, the two cells briefly revolve around each other before escaping and continuing on straight tracks, but at a different angle (\fig \ref{fig:multicell}a, blue). Conversely, in the high adhesion limit, the two crawling cells move directly past one another on their original paths without any angular deflection (\fig \ref{fig:multicell}a, purple).

Deflection, or scattering, was measured in terms of the net angular displacement of the azimuthal angle ($\Delta \phi$) over the period of interaction (\fig \ref{fig:multicell}b). This is motivated by the computational work of \cite{Pooley2007,alexander2008scattering,Farzin2012}, who studied hydrodynamic interactions for pure swimmers. Consistent with our observations in the single-cell case, hydrodynamic scattering effects vanish rapidly with strengthening adhesion, suggesting that strongly-adherent crawling cells can no longer feel each other through the fluid (\fig \ref{fig:multicell}c). Two approaching swimming or weakly-adherent cells interact with each other through perturbations of the fluid, then continue forward along a new, fixed trajectory once they are sufficiently far enough apart to no longer influence each other. By contrast, strongly adherent cells remain on their original paths, unaffected by and seemingly ignorant of the proximity of another cell.

In \fig \ref{fig:multicell} we have assumed that the cells' motion cycles are in phase, but hydrodynamic interactions will also depend on the relative phase between the crawlers' motions (\fig \ref{fig:outphase}). We note that in these cases, the angular displacement $\Delta \phi$ may be a misleading metric for hydrodynamic interactions, as cells can oscillate but remain on their original trajectories. 

The hydrodynamic interactions of the weakly-adherent three-sphere crawlers in \fig \ref{fig:multicell} are similar to those observed for three-sphere swimmers by \cite{Farzin2012}. However, even in the limit of true swimming at zero adhesion, we do not see the large-angle scattering events reported in that paper. We believe this distinction arises from a subtle difference between our numerical methods, indicating that these events may be more dependent on numerical details than immediately apparent.

\subsection*{Crawlers transiently perturb nearby swimmers}\label{sec:res:multikickers}
\begin{figure}[t!]
	\centering
	\includegraphics[width=0.35\textwidth]{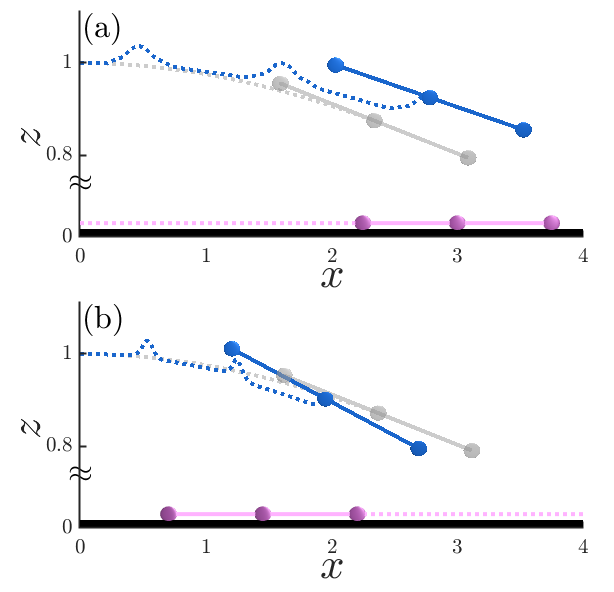}
	\caption{Trajectory snapshots of a swimmer (blue) above a wall under the influence of crawlers (purple, $\xi/6\pi\eta L = 10^4$) moving in the same (a) or opposite (b) directions. Two crawlers have already passed, corresponding to the bumps in the swimmer's path. Lateral dragging from the crawlers causes deviation from the isolated swimmer's path (gray). Sphere radii have been reduced for visualization. Movies are available in the SI. Parameters are listed in Table \ref{table:kicker}.}\label{fig:surfattract}
\end{figure}

Finally, we examine the motion of a cell swimming near a wall and assess its motion under the influence of crawlers on the wall. This is motivated by tumor cell migration and adhesion to blood vessel walls, where the hydrodynamic effects of interstitial flow, matrix geometries, and existing epithelial cells may be significant \cite{Polacheck2011, Pedersen2010}. For these simulations, we choose crawlers and swimmers to have initial directions within the $xz$ plane, allowing for a simpler analysis with the cells remaining in this plane. 

The trajectories of a swimmer in the presence of cells crawling below it are depicted in \fig\ref{fig:surfattract}, with cells crawling in the same direction as the swimmer in a), and opposing the swimmer in b). We see significant deviations from the motion of a swimmer in the absence of crawlers (gray). The swimmer experiences longitudinal bumps and lateral advection in its trajectory corresponding to the passage of a cell crawling underneath. This behavior is consistent with the flow fields shown in \fig\ref{fig:flowfield}--the approaching crawler pushes the fluid up and forward to repel the swimmer but pulls the fluid back down as it passes.

\section*{Discussion}

Our simple three-sphere crawler model makes contact between low-Reynolds number swimming and cell crawling, allowing us to determine the relative prominence of hydrodynamic effects and adhesion-driven motion in adherent cell motility. Sufficiently high adhesion strength ($\xi/6\pi\eta L \gg 1$) will suppress any hydrodynamic effects on a single cell's motion or on two adherent cells, though even strongly adherent cells still generate significant flows around them that can alter the motion of nearby passive particles or swimming cells.  However, depending on which signature of hydrodynamic flow is being observed, the level of adhesion required to suppress it varies. 
For instance, the hydrodynamic drag induced by a substrate significantly reduces a cell's velocity until $\xi/6\pi\eta L \approx 10^1$. Hydrodynamic interactions between cells are expected to be more sensitive to adhesion strength, with suppression observed for adhesion above $\xi/6\pi\eta L \approx 10^{-2}$. 

Our simulations suggest several potential experimental tests for the presence of hydrodynamic effects in crawling cells. First, we note that cells crawling on a fiber will have their motility reduced by the presence of a wall at sufficiently low adhesion strengths (\fig \ref{fig:Substrate}). This effect could be observed in experiments on fibers \cite{sheets2013shape,sharma2013mechanistic}. Secondly, hydrodynamic interactions between weakly adherent cells can be observed (\fig \ref{fig:multicell}). Third, we note that we predict that increasing fluid viscosity can slow the motion of weakly adherent cells (\fig \ref{fig:VCurveStd}). This is in contrast to the limit of freely swimming cells where, holding the shape dynamics constant, changing viscosity will not change swimming speed -- and also the limit of strongly adherent crawlers, where viscosity can be neglected. (Swimmer speed also depends on viscosity when the swimmer's forces, rather than motion, are prescribed, but this arises from a fundamentally different reason \cite{pande2017setting}.) Depending on the experiment, cell type, and viscogen, increased viscosity has been seen to both increase cell speed \cite{gonzalez2018extracellular} and decrease it \cite{matsui2005reduced,folger1978translational}; however, we emphasize that interpreting experiments with increased viscosity can be difficult due to the different effective viscosities at different scales and the effect of external viscosity on receptor dynamics \cite{kobylkevich2018reversing}. We should also note that our plots, such as \fig \ref{fig:VCurveStd}, which describe velocities as a function of $\xi/6\pi \eta L$, show the dependence when $\xi$ is varied, holding $\Fthresh$ constant in simulation units (i.e. holding $\eta = 1$ constant). If $\eta$ is varied, $\Fthresh$ should be constant in real units, not simulation units, and $v_\textrm{cm}$ will not increase with increasing $\eta$.

We see qualitative, but not quantitative, agreement with experiments varying the degree of adhesive coating on the substrate, with our model predicting a slower speed for swimming than for crawling. This is consistent with, e.g., Aoun et al. \cite{Aoun2019}, who see surface-adjacent but nonadherent swimming cells moving with a lower speed than crawling, adherent cells, and the foundational experiments of Barry and Bretscher \cite{Barry2010}. Similarly, calculations by Bae and Bodenschatz demonstrate that, if there is no retrograde cell surface flow, swimming by cell protrusions may be a factor of ten slower than crawling with the same set of shape dynamics \cite{bae2010swimming}. We see a reduction by a factor of around 60 in our model, as we have included fewer details of shape dynamics. However, these results are all broadly consistent with the emerging consensus that the flow of the cell surface is a primary driver of eukaryotic cell swimming \cite{Aoun2019,oneill2018membrane}. As our results do not include membrane flow, we do not expect quantitative agreement. We also note that other mechanisms have been suggested to explain the non-monotonic velocity-adhesion curve, including cell shape changes with adhesive wetting \cite{cao2019cell} and links between adhesivity and protrusion \cite{carlsson2011mechanisms}; we have not addressed either of these aspects. 

Our coarse-grained, minimal model provides intuition for experiments in which the apparent distinction whether cells are swimming or crawling is ambiguous, because hydrodynamic effects may alter a crawling cell's speed. The model suggests that average speed in different conditions, including different viscosities and hydrodynamic geometries, as well as intercellular interactions may be used at least as a qualitative metric to characterize the extent of hydrodynamic effects in motility. Moreover, further improvement and inspection of this model may be able to describe how crawling cells may attract or repel nearby particles or swimmers in the context of problems in collective motility, cancer metastasis, and biofilm dynamics. In particular, we note that \cite{mathijssen2018nutrient} have recently shown that nutrient transport toward the surface can be a consequence of active swimmers near a surface; our results provide a more microscopic view of this problem and how it relates to crawling eukaryotic cells. Extensions of our model could also be made to study mixing induced by eukaryotic cell crawling, as has been done for ciliary carpets \cite{ding2014mixing}. In addition, as the dynamics of swimmers in non-Newtonian and viscoelastic environments has proved to be a fertile area \cite{lauga2011life,elfring2015theory}, it is a natural question what effect these mechanical features will have on crawling cells in biological complex fluids.

\section*{Author Contributions}

MHM developed all the code and carried out all simulations. MHM and BAC designed the research, analyzed data, and wrote the article. 

\section*{Acknowledgments} 
MHM acknowledges support from Johns Hopkins University through the Provost’s Undergraduate Research Award (PURA). We would like to thank Gwynn Elfring for useful comments on a draft of the manuscript, and Yun Chen and Matthew Pittman for valuable conversations and references on viscosity-dependent motility.

\providecommand*{\mcitethebibliography}{\thebibliography}
\csname @ifundefined\endcsname{endmcitethebibliography}
{\let\endmcitethebibliography\endthebibliography}{}


\clearpage
\onecolumn

\beginsupplement

\section*{Supplementary Material}


\appendix

\subsection*{Supplementary Movie Captions}

\begin{itemize}
    \item Movie 1: Anti-parallel cells at low adhesion, corresponding with the parameters of \fig \ref{fig:multicell}a. 
    \item Movie 2: Anti-parallel cells at high adhesion, corresponding with the parameters of \fig \ref{fig:multicell}b.
    \item Movie 3 and 4: Swimmer with four crawlers in the same (3) or opposite (4) direction with a superimposed isolated trajectory. Parameters correspond to those of \fig \ref{fig:surfattract}.
\end{itemize}

\subsection*{Supplementary Figures}

\begin{figure}[h!]
\centering
\includegraphics[width=0.8\textwidth]{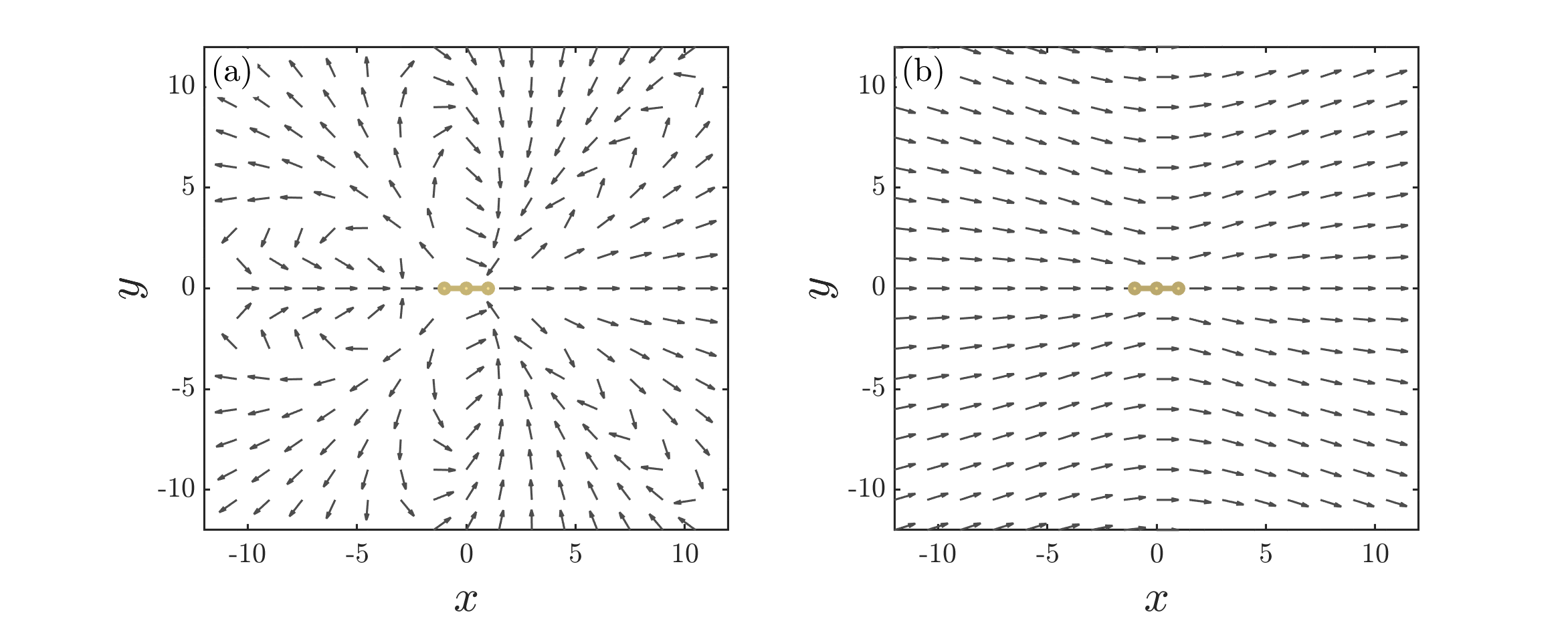}
\caption{Far-field time-averaged flow fields for a swimmer ($\xi/6\pi\eta L = 0$) (a) and a crawler ($\xi/6\pi\eta L = 10^3$) (b) moving to the right far from a wall.}
\label{fig:farfield}
\end{figure}

\begin{figure}[h!]
\centering
\includegraphics[width = 0.95\textwidth]{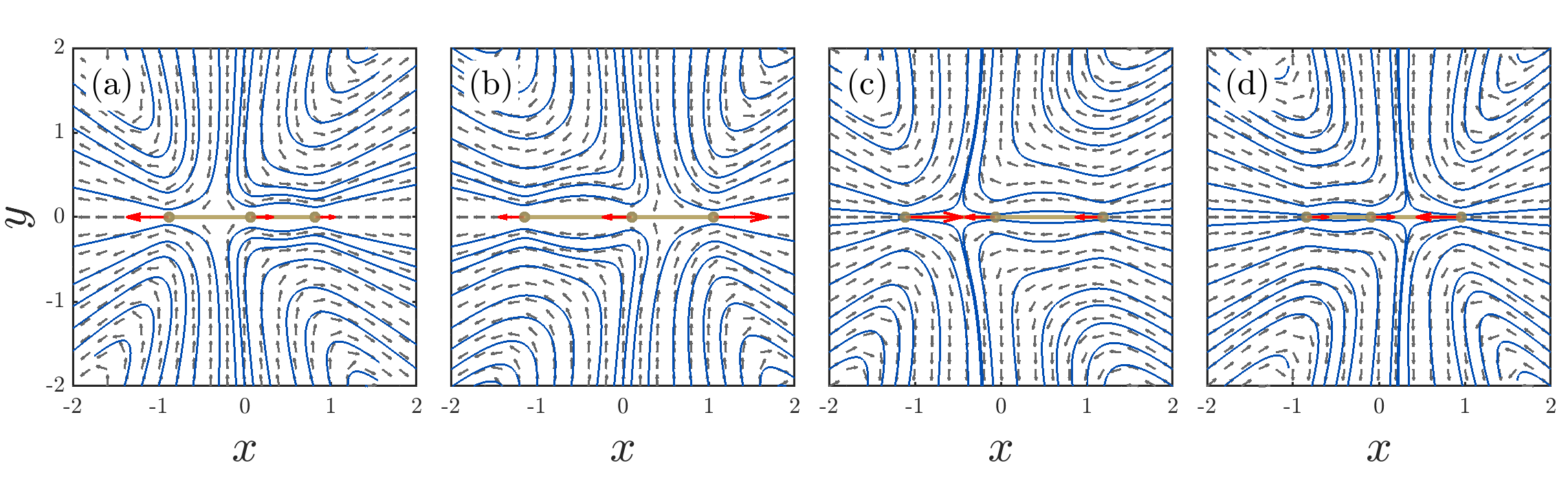}
\caption{Instantaneous streamlines (blue) and flow field (gray) around a swimmer ($\xi/6\pi\eta L = 0$) far from a wall through its axis during each phase of motion. The total force on each bead is illustrated by the red arrows. The flow field around each phase resembles that of a positive (a-b) or negative (c-d) force dipole flow, corresponding to extension or contraction of the cell.}
\label{fig:farfield2}
\end{figure}
\begin{figure}[h!]
\centering
\includegraphics[width = 0.95\textwidth]{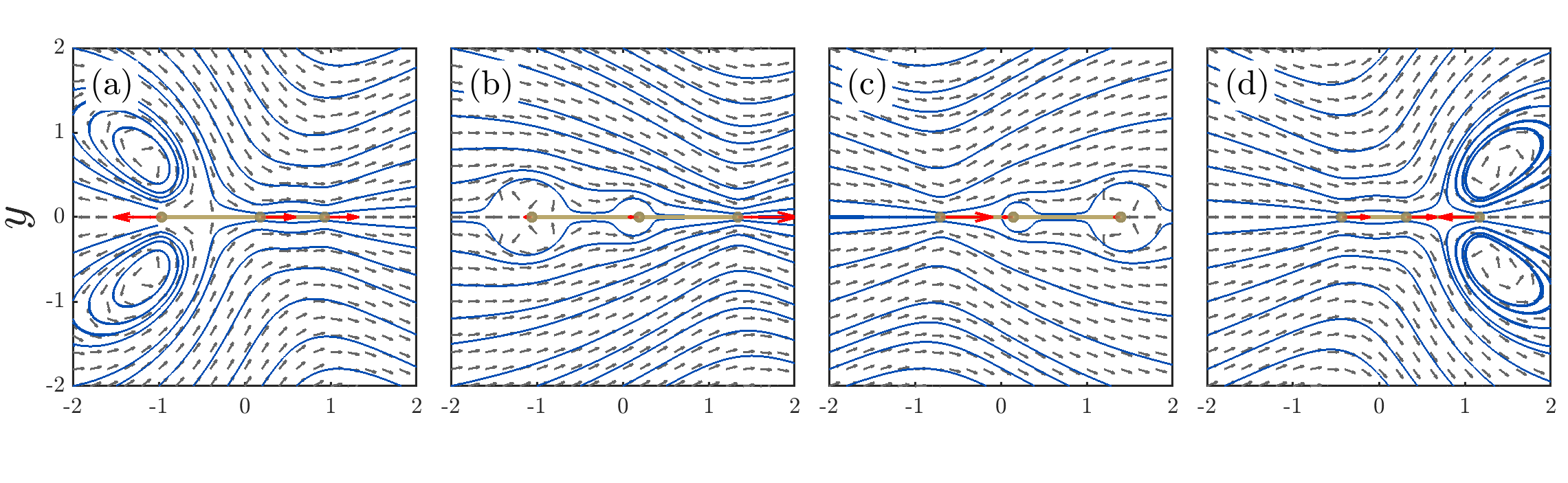}
\caption{Instantaneous streamlines (blue) and flow field (gray) around a crawler ($\xi/6\pi\eta L = 10^3$) on a fiber far from a wall through its axis during each phase of motion. The total force on each bead is illustrated by the red arrows. During trailing arm extension (a), the trailing and center bead must both exert a large internal force to overcome strong adhesion, creating a flow field that roughly resembles a positive force dipole. (b-c) During leading arm extension (b) and trailing arm contraction (c), the beads for the trailing and leading arms, respectively, exert little force due to high adhesion and a zero deformation velocity. As a result, the flow field behaves as that of a Stokeslet around the mobile bead. During leading arm contraction (d), since the leading and center beads are moving in spite of strong adhesion, the flow field again roughly resembles a force dipole, though now negative.}
\label{fig:farfield3}
\end{figure}

\begin{figure}[h!]
\centering
\includegraphics[width = 0.95\textwidth]{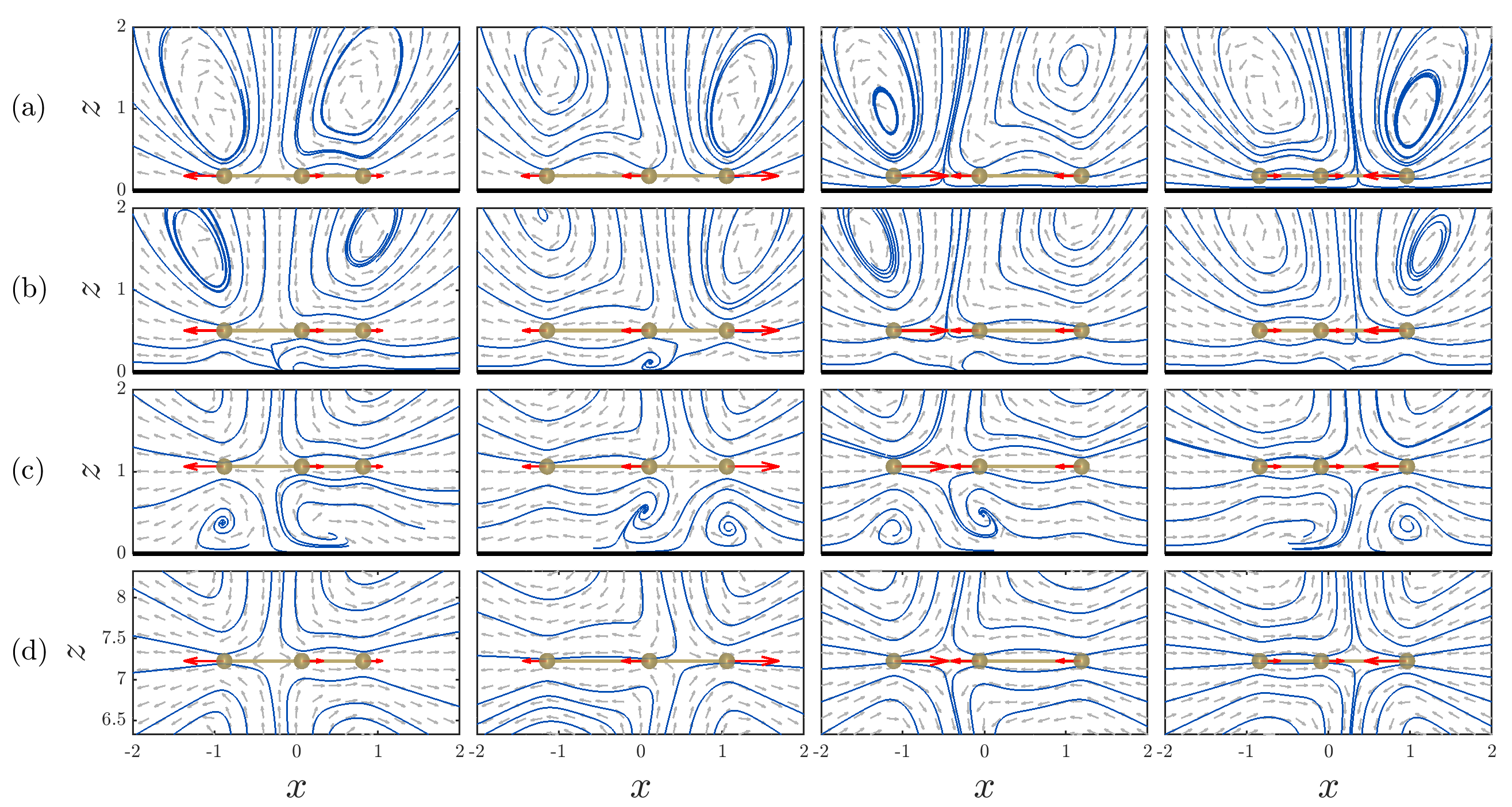}
\caption{Instantaneous streamlines (blue) and flow field (gray) around a cell at very low adhesion ($\xi/6\pi\eta L = 10^{-3}$) on a fiber at varying heights during each phase of motion. The total force on each bead is illustrated by the red arrows. The substrate surface at $z=0$ is shown in the thick black line. (a) $z = 2a$. The cell is just hovering over the surface and is in the first region of increasing velocity in \fig \ref{fig:Substrate}b. (b) $z = 5a$. The cell is high enough off the wall to allow for flow underneath it, but close enough to still be affected by the wall. The cell exists in the intermediate plateau in the velocity profile of \fig \ref{fig:Substrate}b. (c) $z = 10a$. The cell is high enough off the wall to create vortices underneath, and velocity again is increasing. (d) $z > 50a$. The cell is far enough away from the wall to no longer be affected by its hydrodynamics. Note that in these figures, as in \fig \ref{fig:flowfield} above, local streamlines can be misleading; no-slip boundary conditions are obeyed at z = 0.} 
\label{fig:wall_flows}
\end{figure}

\begin{figure}[h!]
\centering
\includegraphics[width=\textwidth]{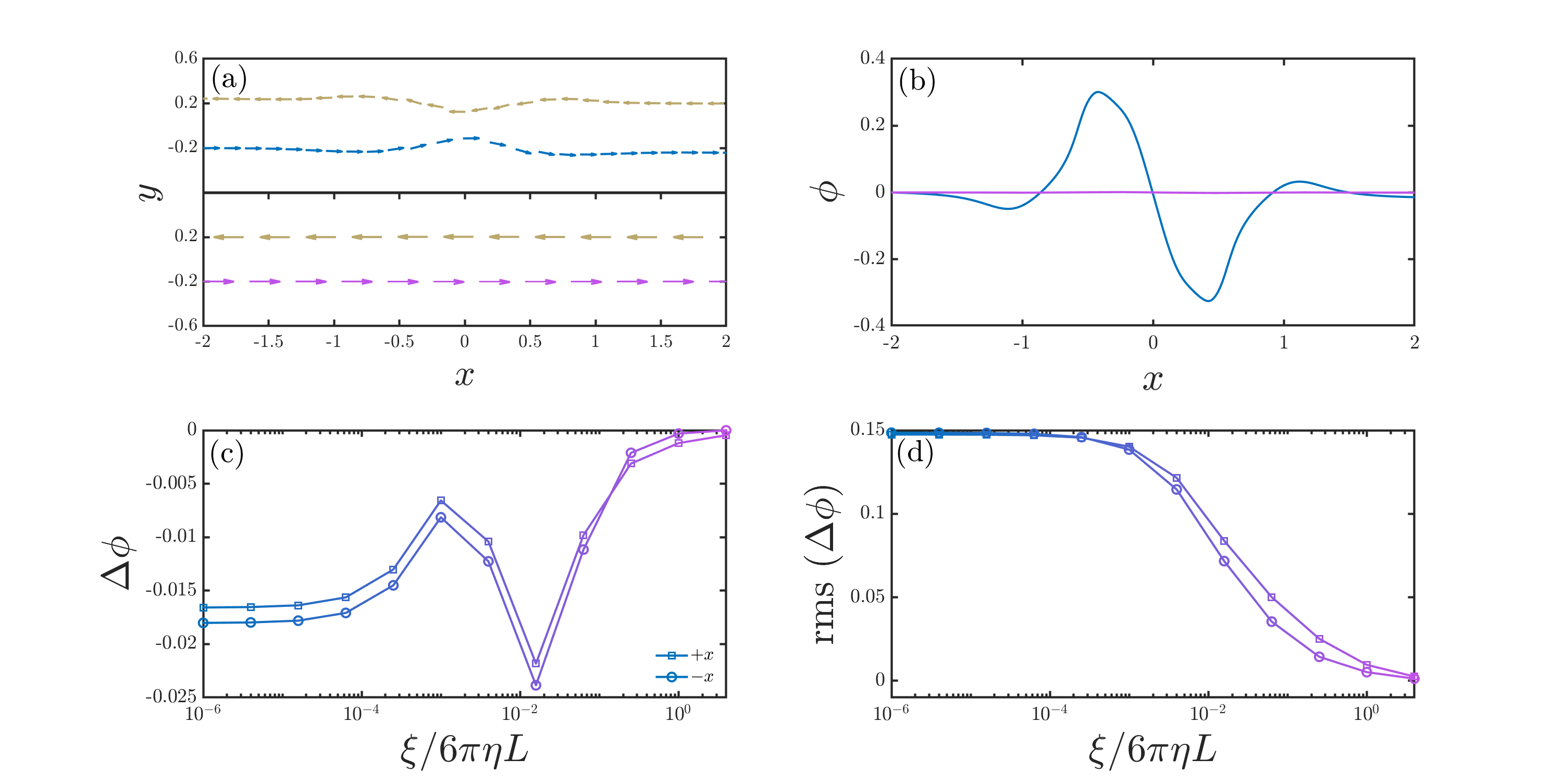}
\caption{(a) Trajectories for two antiparallel cells, out of phase by half a motion cycle, at low (blue) and high (purple) adhesion. The $-x$ oriented cell is shown in yellow. (b) Angle $\phi$ of the $+x$-oriented cell, shown as a function of the cell's center of mass position (large $x$ corresponding to post-interaction). (c) Net angular deflection ($\Delta\phi$) of the $+x$ (square) and $-x$ (circle) oriented cells as a function of $\xi/6\pi\eta L$ is both small and non-monotonic with adhesion strength, suggesting that sometimes another metric must be used to characterize hydrodynamic interactions. (d) RMS deflection, calculated per cycle over the period of interaction, defined as the time over which the cells are within a certain distance of each other (in this case $d = 4L$), exhibits monotonic behavior over $\xi/6\pi\eta L$ and may be a useful metric of hydrodynamic interactions when net deflection is insufficient.}
\label{fig:outphase}
\end{figure}

\clearpage

\section{Analytical results in the large-adhesion limit}
\label{app:analytic} 

In the large-adhesion limit, we can neglect hydrodynamics and \eq \ref{eq:VMFfinal} is equivalent to:
\begin{equation}
    \vb{V} = \frac{1}{\boldsymbol{\xi}(t)} \circ \Fint,
\end{equation}
i.e. the motion of one bead is only controlled by the force on that bead and the friction coefficient on the bead. This allows us to simply compute the velocity of a single cell in the high-adhesion limit. To do this, we simplify to one dimension, following the approach of \cite{Golestanian2008}, and find the internal forces that satisfy
\begin{align}
V_3-V_2 = F_3^\textrm{int}/\xi_3 - F_2^\textrm{int}/\xi_2 &= W_L \\
V_2-V_1 = F_2^\textrm{int}/\xi_2 - F_1^\textrm{int}/\xi_1 &= W_T \\
F_1^\textrm{int} + F_2^\textrm{int} + F_3^\textrm{int} &= 0.
\end{align}
This can be done analytically due to the simplicity of the model, though we do not write it explicitly here. These forces then determine
\begin{equation}
v_\textrm{cm}(t) = \frac{1}{3}(V_1+V_2+V_3) = \frac{-2 W_L \mu_1 \mu_2 + 2 W_T \mu_2 \mu_3 + W_L(\mu_1+\mu_2)\mu_3 - W_T \mu_1(\mu_2 + \mu_3)}{3 \mu_2 \mu_3 + 3 \mu_1 (\mu_2 + \mu_3)}
\end{equation}
where $\mu_i = 1/\xi_i$ is a mobility for bead $i$. This gives the center-of-mass velocity at a given instant, and depends on $W_L$ and $W_T$ as well as $\xi_i(t)$ for each bead. In addition, the forces determine the maximum required force $\textrm{max} |\Freq|$; we scale the internal forces as in the main text if $\textrm{max} |\Freq|$ exceeds \Fthresh. 

We can then compute the time average of $v_\textrm{cm}(t)$ over one whole cycle. During each phase of the cycle, the center of mass velocity is constant, so this time average is merely
\begin{align}
    v_\textrm{cm} = \frac{1}{T_\textrm{trail-ext} + T_\textrm{lead-ext} + T_\textrm{trail-cont} + T_\textrm{lead-cont}} &\times \\ \nonumber [T_\textrm{trail-ext} v_{cm}^\textrm{trail-ext} + T_\textrm{lead-ext} v_{cm}^\textrm{lead-ext} + T_\textrm{trail-cont} &v_{cm}^\textrm{trail-cont} + T_\textrm{lead-cont} v_{cm}^\textrm{lead-cont} ]
\end{align}
where the velocities for each phase are worked out by choosing the appropriate values of $\xi_1$, $\xi_2$, and $\xi_3$ and $W_L$ and $W_T$ from Table \ref{table:xivals}. Note that for working out the time $T$ of each phase, if the arm is contracting with a constant rate $W$, this time is merely $\Delta L / W$; however, if the force is above the threshold, then the contraction will be slower, taking a time $\frac{\Delta L}{W}\times\frac{\textrm{max} |\Freq|}{\Fthresh}$. Computing the average, and simplifying, yields \eq \ref{eq:analytic} in the main text. We have found computer algebra systems ({\it Mathematica}) useful for keeping track of the special cases for when the force exceeds the threshold. 

\begin{figure}[h!]
\centering
\includegraphics[width=0.6\textwidth]{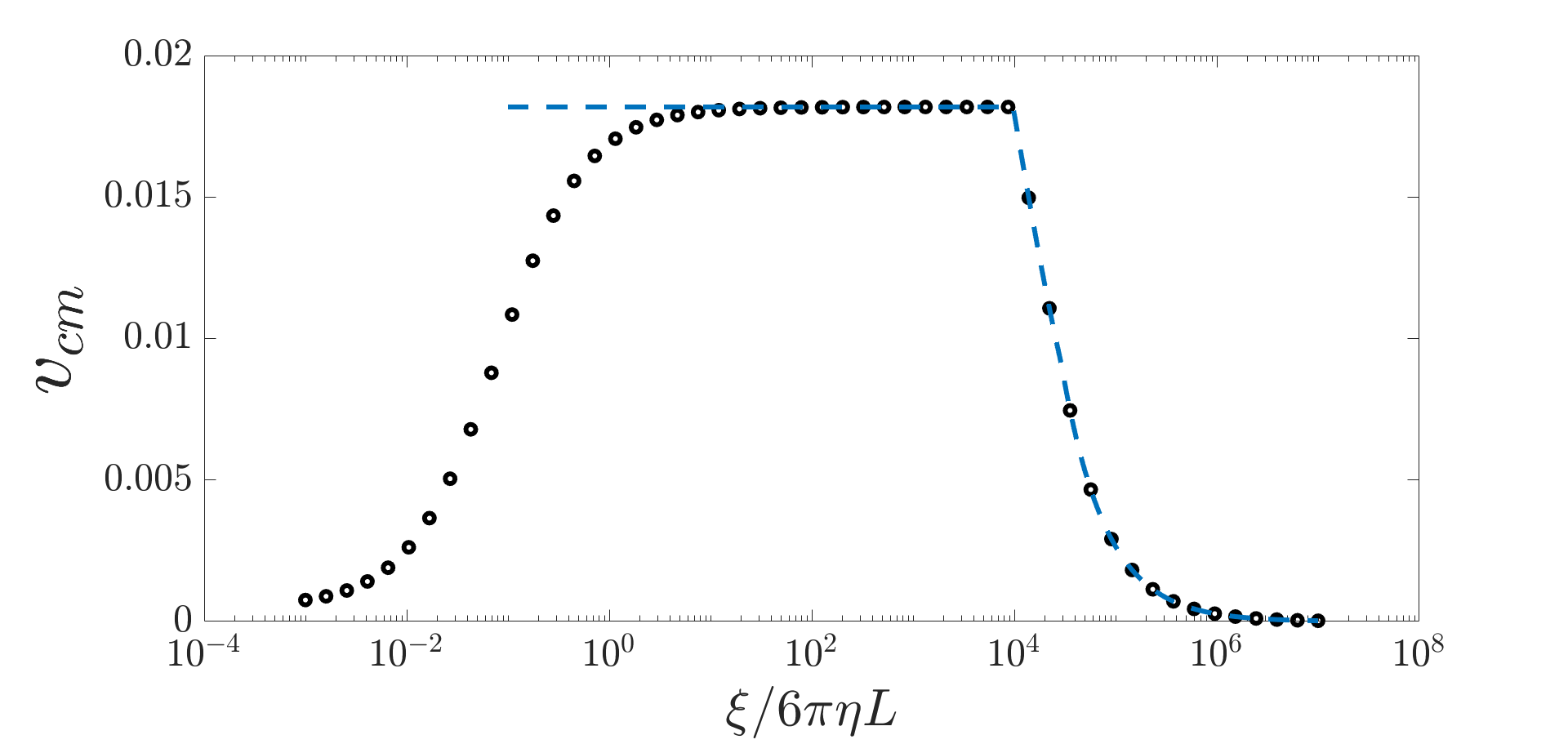}
\caption{Comparison between full simulation and the high-friction asymptotic result of \eq \ref{eq:analytic}. $\Fthresh = 10^4$; all other parameters are as in Table \ref{table:stdpar}.}
\label{fig:asymptotic}
\end{figure}

We find that this analytic result captures our full simulations very well in the large-adhesion limit (\fig \ref{fig:asymptotic}).

\section{The regularized Blake tensor}\label{app:BlakeTensor}

We apply the results of \cite{Ainley2008} to compute how the fluid velocity at point $\vb{R}_m$ depends on a force $\vb{F}_n$ exerted at point $\vb{R}_n$, in the presence of a substrate with a no-slip boundary condition at $z = 0$ (with a normal vector of $\hat{e}_z$). This regularized Green's function generalizes the simple expressions given in the main paper (\eq \ref{eq:oseen} and \eq \ref{eq:beadbead}). The result of \cite{Ainley2008} is:
\begin{align*}
\vb{V}(\vb{R}_m, \vb{R}_n) &= [\vb{F}_nH_1(r_n^*) + (\vb{F}_n\cdot\vb{r}_n^*)\vb{r}_n^*H_2(r_n^*)]\\
&\hspace{0.3cm}- [\vb{F}_nH_1(r_n) + (\vb{F}_n \cdot \vb{r}_n)\vb{r}_nH_2(r_n)]\\
&\hspace{0.3cm}- h_n^2[\vb{g}_nD_1(r_n) + (\vb{g}_n\cdot\vb{r}_n)\vb{r}_nD_2(r_n)]\\
&\hspace{0.3cm} + 2h_n\left[\frac{H_1'(r_n)}{r_n}+H_2(r_n)\right]\left[(\vb{F}_n\times \hat{e}_z)\times\vb{r}_n\right]\\
&\hspace{0.3cm}+ 2h_n\left[\left(\vb{r}_ng_{nz} + \vb{g}_nz_n\right)H_2(r_n) + (\vb{g}_n\cdot\vb{r}_n)\left(\hat{e}_z\frac{H_1'(r_n)}{r_n}+\textit{}\vb{r}_nz_n\frac{H_2'(r_n)}{r_n}\right)\right]\numberthis\label{eq:RegTensorOrig}
\end{align*}
We have defined two relative distances: first, the explicit distance coordinate between the two points, $\vb{r}_n^* = \vb{R}_m - \vb{R}_n$, and the image distance coordinate between the target sphere and the image of the force-generating sphere, $\vb{r}_n = \vb{R}_m - \vb{R}_n + 2h_n\hat{e}_z$. Here, $h$ is the distance of a sphere from the surface, $\vb{g}_n = 2(\vb{F}_n\cdot\hat{e}_z)\hat{e}_z - \vb{F}_n = (-F_{nx}, -F_{ny}, F_{nz})^T$ and $z_n$ is the $z$-component of $\vb{r}_n$. Note that $z_n$ refers to the relative coordinate $\vb{r}_n$ and is not equivalent to $h_n$, which is the $z$-component of the absolute coordinate $\vb{R}_n$. The full derivation for \eq \ref{eq:RegTensorOrig} is available in Reference \cite{Ainley2008}, but we note that the expression above corrects a sign error in its fourth bracketed term, which refers to the image rotlets.

\eq \ref{eq:RegTensorOrig} is still well-defined when $R_m = R_n$. The forces are smoothed over the sphere volumes using four scalar regularization, or "blob," functions:
\begin{align}
H_1(r) &= \frac{1}{8\pi(r^2 + \epsilon^2)^{1/2}} + \frac{\epsilon^2}{8\pi(r^2+\epsilon^2)^{3/2}}\\
H_2(r) &= \frac{1}{8\pi(r^2+\epsilon^2)^{3/2}}\\
D_1(r) &= \frac{1}{4\pi(r^2 + \epsilon^2)^{3/2}} - \frac{3\epsilon^2}{4\pi(r^2+\epsilon^2)^{5/2}}\\
D_2(r) &= - \frac{3}{4\pi(r^2+\epsilon^2)^{5/2}}
\end{align}
Here $\epsilon$ defines the width of these functions and, therefore, the characteristic length over which to smooth the forces. Thus we set $\epsilon$ equal to the sphere radius $a$ so that the sphere becomes a ball of finite force density instead of a singular point force. 

The form \eq \ref{eq:RegTensorOrig} is useful only if the forces on each sphere are already known. Since we must also calculate the individual forces in addition to the velocities through \eq \ref{eq:VMFfull}, the expression in \eq \ref{eq:RegTensorOrig} can be rearranged into a more functional form, which is a 3$\times$3 mobility submatrix defined by the hydrodynamic interaction from sphere $n$ acting on sphere $m$, $\hat{s}_{n\rightarrow m}$, that satisfies the relationship defined in \eq \ref{eq:beadbead}. We define the mobility submatrices: 
\begin{align*}
\eta\hat{s}_{n\rightarrow m} &= 2H_2(r)\begin{pmatrix}
-2hz & 0 & x(h - z)\\
0 & 0 & y(h-z)\\
hx & hy & z(2h-z)
\end{pmatrix} 
+ H_2(r^*)\begin{pmatrix}
x^{*2} & x^*y^* & x^*z^*\\
x^*y^* & y^{*2} & y^*z^*\\
x^*z^* & y^*z^* & z^{*2}
\end{pmatrix} \\
&\hspace{0.5cm} + \left[h^2D_2(r) - H_2(r) - 2h\left(\frac{H_2'(r)}{r}\right)z\right]\begin{pmatrix}
x^2 & xy & -xz\\
xy & y^2 & -yz\\
xz & yz & -z^2
\end{pmatrix} \\
&\hspace{0.5cm} 
+ h^2D_1(r)\begin{pmatrix}
1 & 0 & 0\\
0 & 1 & 0\\
0 & 0 & -1
\end{pmatrix} + 2h\left(\frac{H_1'(r)}{r}\right)z\begin{pmatrix}
-1 & 0 & 0\\
0 & 1 & 0\\
0 & 0 & 1
\end{pmatrix}\\
&\hspace{0.5cm}+ \left(H_1(r^*) - H_1(r)\right)\mathbb{I}\numberthis\label{eq:Smatmn}
\end{align*}
in which the subscripted $n$ is implied on all relevant terms. The standard mobility matrix, $\hat{M}$, is then an arrangement of these mobility submatrices:
\begin{equation}
\hat{M}= \begin{pmatrix}
\hat{s}_{1\rightarrow 1} & \ldots & \hat{s}_{N\rightarrow 1}\\
\vdots & \ddots & \vdots\\
\hat{s}_{1\rightarrow N} & \ldots & \hat{s}_{N\rightarrow N}
\end{pmatrix}\label{eq:MfromS}
\end{equation}
To solve the system with friction given by \eq \ref{eq:VMFfinal}, the modified mobility matrix, $\hat{\mathcal{M}}$, is numerically calculated. The submatrices, $\hat{\mathcal{S}}$, are then defined such that \begin{equation}
\left(\mathbb{1} + \hat{\Xi}\right)^{-1}\hat{M} = \hat{\mathcal{M}} = \begin{pmatrix}
\hat{\mathcal{S}}_{1\rightarrow 1} & \ldots & \hat{\mathcal{S}}_{N\rightarrow 1}\\
\vdots & \ddots & \vdots\\
\hat{\mathcal{S}}_{1\rightarrow N} & \ldots & \hat{\mathcal{S}}_{N\rightarrow N}
\end{pmatrix}.\label{eq:modifiedS}
\end{equation}
which can now be used for all calculations. For example, to rewrite the constraints presented in \eq \ref{cst:WL}-\ref{cst:gamma}, 
\begin{align}
W_L &= \left[\sum_n \left(\hat{\mathcal{S}}_{n\rightarrow 3} - \hat{\mathcal{S}}_{n\rightarrow 2}\right)\vb{F}_n\right]\cdot\hat{\alpha}\label{eq:WLc}\\
W_T &= \left[\sum_n \left(\hat{\mathcal{S}}_{n\rightarrow 2} - \hat{\mathcal{S}}_{n\rightarrow 1}\right)\vb{F}_n\right]\cdot\hat{\alpha}\label{eq:WTc}\\
0 &= \biggl\lbrace\sum_n \left[L_T\left(\hat{\mathcal{S}}_{n \rightarrow 3} - \hat{\mathcal{S}}_{n\rightarrow 2}\right) - L_L\left(\hat{\mathcal{S}}_{n\rightarrow 2} - \hat{\mathcal{S}}_{n\rightarrow 1}\right) \right]\vb{F}_n\biggr\rbrace \cdot \hat{\beta}\label{eq:betac}\\
0 &= \biggl\lbrace\sum_n \left[L_T\left(\hat{\mathcal{S}}_{n \rightarrow 3} - \hat{\mathcal{S}}_{n\rightarrow 2}\right) - L_L\left(\hat{\mathcal{S}}_{n\rightarrow 2} - \hat{\mathcal{S}}_{n\rightarrow 1}\right) \right]\vb{F}_n \biggr\rbrace\cdot \hat{\gamma}\label{eq:gammac}
\end{align}
\clearpage

\section{Constraint matrices} \label{app:cstmtx}
The constraint matrices are explicitly defined here, using the same notation as before, and where $\delta L = L_T - L_L$.
\begin{center}
	\rotatebox{-90}{\begin{minipage}{0.92\textheight}
{\small 	\begin{align*}
	&\hat{C}_{swim}\vb{d}_{swim} = \\
	&\begin{pmatrix}
	\begin{tabularx}{7.5in}{Y *{8}{Y}}
	1 & 0 & 0 & 1 & 0 & 0 & 1 & 0 & 0\\
	0 & 1 & 0 & 0 & 1 & 0 & 0 & 1 & 0\\
	0 & 0 & 1 & 0 & 0 & 1 & 0 & 0 & 1\\
    $ \mathcal{T}_{Li1}\beta_i $ &  $ \mathcal{T}_{Li2}\beta_i $ & $ \mathcal{T}_{Li3}\beta_i $ &$  0 $ & $ 0 $ & $ 0 $ &$  \mathcal{T}_{Ti1}\beta_i $ & $ \mathcal{T}_{Ti2}\beta_i $ & $ \mathcal{T}_{Ti3}\beta_i $\\
	$ \mathcal{T}_{Li1}\gamma_i $ &  $ \mathcal{T}_{Li2}\gamma_i $ & $ \mathcal{T}_{Li3}\gamma_i $ &$  0 $ & $ 0 $ & $ 0 $ &$  \mathcal{T}_{Ti1}\gamma_i $ & $ \mathcal{T}_{Ti2}\gamma_i $ & $ \mathcal{T}_{Ti3}\gamma_i $\\
	\multicolumn{3}{c}{$\left(\hat{\mathcal{S}}_{1\rightarrow 3} - \hat{\mathcal{S}}_{1\rightarrow 2}\right)\cdot\hat{\alpha}$} & \multicolumn{3}{c}{$\left(\hat{\mathcal{S}}_{2\rightarrow 3} - \hat{\mathcal{S}}_{2\rightarrow 2}\right)\cdot\hat{\alpha}$} & \multicolumn{3}{c}{$\left(\hat{\mathcal{S}}_{3\rightarrow 3} - \hat{\mathcal{S}}_{3\rightarrow 2}\right)\cdot\hat{\alpha}$}\\
	\multicolumn{3}{c}{$\left(\hat{\mathcal{S}}_{1\rightarrow 2} - \hat{\mathcal{S}}_{1\rightarrow 1}\right)\cdot\hat{\alpha}$} & \multicolumn{3}{c}{$\left(\hat{\mathcal{S}}_{2\rightarrow 2} - \hat{\mathcal{S}}_{2\rightarrow 1}\right)\cdot\hat{\alpha}$} & \multicolumn{3}{c}{$\left(\hat{\mathcal{S}}_{3\rightarrow 2} - \hat{\mathcal{S}}_{3\rightarrow 1}\right)\cdot\hat{\alpha}$}\\
	\multicolumn{3}{c}{$\left[L_T\hat{\mathcal{S}}_{1 \rightarrow 3} + L_L\hat{\mathcal{S}}_{1\rightarrow 1} - \delta L\hat{\mathcal{S}}_{1\rightarrow 2} \right] \cdot \hat{\beta}$} & \multicolumn{3}{c}{$\left[L_T\hat{\mathcal{S}}_{2 \rightarrow 3} + L_L\hat{\mathcal{S}}_{2\rightarrow 1} - \delta L\hat{\mathcal{S}}_{2\rightarrow 2} \right] \cdot \hat{\beta}$} & \multicolumn{3}{c}{$\left[L_T\hat{\mathcal{S}}_{3 \rightarrow 3} + L_L\hat{\mathcal{S}}_{3\rightarrow 1} - \delta L\hat{\mathcal{S}}_{3\rightarrow 2} \right] \cdot \hat{\beta}$} \\
	\multicolumn{3}{c}{$\left[L_T\hat{\mathcal{S}}_{1 \rightarrow 3} + L_L\hat{\mathcal{S}}_{1\rightarrow 1} - \delta L\hat{\mathcal{S}}_{1\rightarrow 2} \right] \cdot \hat{\gamma}$} & \multicolumn{3}{c}{$\left[L_T\hat{\mathcal{S}}_{2 \rightarrow 3} + L_L\hat{\mathcal{S}}_{2\rightarrow 1} - \delta L\hat{\mathcal{S}}_{2\rightarrow 2} \right] \cdot \hat{\gamma}$} & \multicolumn{3}{c}{$\left[L_T\hat{\mathcal{S}}_{3 \rightarrow 3} + L_L\hat{\mathcal{S}}_{3\rightarrow 1} - \delta L\hat{\mathcal{S}}_{3\rightarrow 2} \right] \cdot \hat{\gamma}$} 
	\end{tabularx}
	\end{pmatrix}\begin{pmatrix}
	0\\0\\0\\0\\0\\W_L\\W_T\\0\\0
	\end{pmatrix}\begin{matrix}
	i\\ii\\iii\\iv\\v\\vi\\vii\\viii\\ix
	\end{matrix}\numberthis\label{eq:Cswim}
	\end{align*}
	\begin{align*}
	&\hat{C}_{crawl}\vb{d}_{crawl} = \\
	&\begin{pmatrix}
	\begin{tabularx}{7in}{Y *{8}{Y}}
	1 & 0 & 0 & 1 & 0 & 0 & 1 & 0 & 0\\
	0 & 1 & 0 & 0 & 1 & 0 & 0 & 1 & 0\\
	$ \mathcal{T}_{Li1}\beta_i $ &  $ \mathcal{T}_{Li2}\beta_i $ & $ \mathcal{T}_{Li3}\beta_i $ &$  0 $ & $ 0 $ & $ 0 $ &$  \mathcal{T}_{Ti1}\beta_i $ & $ \mathcal{T}_{Ti2}\beta_i $ & $ \mathcal{T}_{Ti3}\beta_i $\\
	\multicolumn{3}{c}{$\left(\hat{\mathcal{S}}_{1\rightarrow 3} - \hat{\mathcal{S}}_{1\rightarrow 2}\right)\cdot\hat{\alpha}$} & \multicolumn{3}{c}{$\left(\hat{\mathcal{S}}_{2\rightarrow 3} - \hat{\mathcal{S}}_{2\rightarrow 2}\right)\cdot\hat{\alpha}$} & \multicolumn{3}{c}{$\left(\hat{\mathcal{S}}_{3\rightarrow 3} - \hat{\mathcal{S}}_{3\rightarrow 2}\right)\cdot\hat{\alpha}$}\\
	\multicolumn{3}{c}{$\left(\hat{\mathcal{S}}_{1\rightarrow 2} - \hat{\mathcal{S}}_{1\rightarrow 1}\right)\cdot\hat{\alpha}$} & \multicolumn{3}{c}{$\left(\hat{\mathcal{S}}_{2\rightarrow 2} - \hat{\mathcal{S}}_{2\rightarrow 1}\right)\cdot\hat{\alpha}$} & \multicolumn{3}{c}{$\left(\hat{\mathcal{S}}_{3\rightarrow 2} - \hat{\mathcal{S}}_{3\rightarrow 1}\right)\cdot\hat{\alpha}$}\\
	\multicolumn{3}{c}{$\left[L_T\hat{\mathcal{S}}_{1 \rightarrow 3} + L_L\hat{\mathcal{S}}_{1\rightarrow 1} - \delta L\hat{\mathcal{S}}_{1\rightarrow 2} \right] \cdot \hat{\gamma}$} & \multicolumn{3}{c}{$\left[L_T\hat{\mathcal{S}}_{2 \rightarrow 3} + L_L\hat{\mathcal{S}}_{2\rightarrow 1} - \delta L\hat{\mathcal{S}}_{2\rightarrow 2} \right] \cdot \hat{\gamma}$} & \multicolumn{3}{c}{$\left[L_T\hat{\mathcal{S}}_{3 \rightarrow 3} + L_L\hat{\mathcal{S}}_{3\rightarrow 1} - \delta L\hat{\mathcal{S}}_{3\rightarrow 2} \right] \cdot \hat{\gamma}$} \\
	\multicolumn{3}{c}{$\hat{\mathcal{S}}_{1\rightarrow 1}\cdot\hat{e}_z$} & \multicolumn{3}{c}{$\hat{\mathcal{S}}_{2\rightarrow 1}\cdot\hat{e}_z$} & \multicolumn{3}{c}{$\hat{\mathcal{S}}_{3\rightarrow 1}\cdot\hat{e}_z$} \\
	\multicolumn{3}{c}{$\hat{\mathcal{S}}_{1\rightarrow 2}\cdot\hat{e}_z$} & \multicolumn{3}{c}{$\hat{\mathcal{S}}_{2\rightarrow 2}\cdot\hat{e}_z$} & \multicolumn{3}{c}{$\hat{\mathcal{S}}_{3\rightarrow 2}\cdot\hat{e}_z$} \\
	\multicolumn{3}{c}{$\hat{\mathcal{S}}_{1\rightarrow 3}\cdot\hat{e}_z$} & \multicolumn{3}{c}{$\hat{\mathcal{S}}_{2\rightarrow 3}\cdot\hat{e}_z$} & \multicolumn{3}{c}{$\hat{\mathcal{S}}_{3\rightarrow 3}\cdot\hat{e}_z$}
	\end{tabularx}
	\end{pmatrix}\begin{pmatrix}
	0\\0\\0\\W_L\\W_T\\0\\0\\0\\0
	\end{pmatrix}\begin{matrix}
	i\\ii\\iii\\iv\\v\\vi\\vii\\viii\\ix
	\end{matrix}\numberthis\label{eq:Ccrawl}
	\end{align*}}
\end{minipage}}
\end{center}

When we define the constraints that keep the cell torque-free, we use  
\begin{align*}
\hat{\mathcal{T}}_L &= \begin{pmatrix}
0 & z_L & -y_L\\
-z_L & 0 & x_L\\
y_L & -x_L & 0
\end{pmatrix} & 
\hat{\mathcal{T}}_T &= \begin{pmatrix}
0 & z_T & -y_T\\
-z_T & 0 & x_T\\
y_T & -x_T & 0
\end{pmatrix}
\end{align*}
where $x_L$ indicates the displacement of the leading arm in the $x$ direction, etc.

The rows of each constraint matrix correspond to the following constraints:

For swimming (\eq \ref{eq:Cswim}), 
\begin{enumerate}
	\item [i-iii:] Force-free conditions
	\item [iv-v:] Torque-free conditions (projection onto $\hat{\beta}$ and $\hat{\gamma}$)
	\item [vi:] Leading arm deformation velocity
	\item [vii:] Trailing arm deformation velocity
	\item [viii-ix:] Rigid body conditions with projections of the deformation velocities onto $\hat{\beta}$ and $\hat{\gamma}$, respectively
\end{enumerate}
For crawling (\eq \ref{eq:Ccrawl}), 
\begin{enumerate}
	\item [i-ii:] Force-free conditions
	\item [iii:] $z$-component ($\hat{\beta}$-projection) of the torque-free condition
	\item [iv:] Leading arm deformation velocity
	\item [v:] Trailing arm deformation velocity
	\item [vi:] Rigid body condition with projection of the deformation velocities onto $\hat{\gamma}$
	\item [vii-ix:] Zero $z$-directional velocities for each sphere
\end{enumerate}
\clearpage

\section{Parameters}\label{app:params}
Below are the parameters used for the simulations discussed in this project. Parameters are given in the simulation units discussed in the Methods of the main paper. Table \ref{table:stdpar} provides all the parameters for the standard simulation. Tables \ref{table:antipar}-\ref{table:kicker} refer to their respective simulations discussed above. Parameters not listed in Tables \ref{table:antipar}-\ref{table:kicker} are unchanged from the standard parameter values given in Table \ref{table:stdpar}.

\begin{table}[h!]
	\caption{Parameters for the standard simulation}
	\centering
	\begin{tabular*}{0.6\textwidth}{c @{\extracolsep{\fill}} c c}
		\hline
		Parameter & Symbol & Value\\
		\hline
		Mean arm length & $L$ & 1\\
		Bead radius & $a$ & 0.1\\
		Initial center of mass & $\vb{R}_{com}(t=0)$ & $(0, 0, a)$\\
		Deformation magnitude & $\Delta L$ & $\pm 0.5$\\
		Deformation velocities & $W_L^+, W_L^-, W_T^+, W_T^-$ & $\pm 0.1$\\
		Polar angle & $\theta$ & $\pi/2$\\
		Azimuthal angle & $\phi$ & 0\\
		High adhesion scale & $\xi_{high}$ & $\xi$\\
		Low adhesion scale & $\xi_{low}$ & 0.2$\xi$\\
		Viscosity & $\eta$ & 1\\
		Threshold force & $\Fthresh$ & $10^3$\\
		Time step & $\Delta t$ & $10^{-2}$\\
		\hline
	\end{tabular*}\label{table:stdpar}
\end{table}

\begin{table}[h!]
	\caption{Parameters for the anti-aligned pair}
	\centering
	\begin{tabular*}{0.6\textwidth}{c @{\extracolsep{\fill}} c c}
		\hline
		Parameter & Cell 1 & Cell 2\\
		\hline
		Initial center of mass, $\vb{R}_{com}(t=0)$ & (-2, -0.2, 0.1) & (2, 0.2, 0.1)\\
		Azimuthal Angle, $\phi$ & 0 & $\pi$ \\
		Global adhesion (swim), $\xi$ & \multicolumn{2}{c}{10$^{-6}\cdot 6\pi$}\\
		Global adhesion (crawl), $\xi$ & \multicolumn{2}{c}{10$^3\cdot 6\pi$}\\
		\hline
	\end{tabular*}\label{table:antipar}
\end{table}

\begin{table}[h!]
\caption{Parameters for the swimmer among multiple crawlers. Crawlers are generated every 90 cycles and removed from the system once the distance between the swimmer and crawler is sufficiently large (after around 60 cycles) to reduce computational load. $\vb{R}_{com}^{c, \pm x}(t_{gen})$ refers to the center of mass of each crawler at the time of its generation.}

\centering
\begin{tabular*}{0.6\textwidth}{c @{\extracolsep{\fill}} c}
	\hline
	Parameter & Value \\
	\hline
	$\vb{R}_{com}^s(t=0)$ & (0, 0, 1)\\
	Time between crawlers & 90 cycles\\
	$\vb{R}_{com}^{c, +x}(t_{gen})$ & (-3, 0, 0.1) \\
	$\vb{R}_{com}^{c, -x}(t_{gen})$ & (8, 0, 0.1) \\
	Crawler deformation velocity, $W_{c}$ & $\pm 0.05$\\
	Crawler deformation magnitude $\Delta L_{c}$ & $\pm0.25$\\
	Azimuthal Angle, $\phi$ & 0 or $\pi$ \\
	Global adhesion (swim), $\xi$ & 0\\
	Global adhesion (crawl), $\xi$ & 10$^4\cdot 30\pi$\\
	\Fthresh & $10^5$\\
	\hline
\end{tabular*}\label{table:kicker}
\end{table}

\clearpage

\end{document}